\documentclass[twocolumn]{aastex631}
\usepackage{graphicx }
\usepackage{lineno}
\usepackage{float}
\usepackage[caption=false]{subfig}
\usepackage{longtable}

\begin{document}

\title{Multi-wavelength observations of substructures in solar flare ribbons}

\author[0000-0002-9911-2285]{Vishal Singh}
\affiliation{Northumbria University, Newcastle upon Tyne, NE1 8ST, UK}

\author[0000-0001-9590-6427]{Eamon Scullion}
\affiliation{Northumbria University, Newcastle upon Tyne, NE1 8ST, UK}

\author[0000-0002-5915-697X]{Gert J. J. Botha}
\affiliation{Northumbria University, Newcastle upon Tyne, NE1 8ST, UK}

\author[0000-0001-6583-1989]{Natasha L. S. Jeffrey}
\affiliation{Northumbria University, Newcastle upon Tyne, NE1 8ST, UK}

\author[0000-0001-9845-266X]{Malcolm Druett}
\affiliation{Plasma Dynamics Group, School of Electrical and Electronic Engineering, University of Sheffield, Sheffield, S1 3JD, UK}

\author[0000-0002-0484-7634]{Alexander G. M. Pietrow }
\affiliation{Leibniz-Institut f\"ur Astrophysik Potsdam (AIP), An der Sternwarte 16, 14482 Potsdam,
Germany}
\author{Aidan O'Flannagain}
\affiliation{Trinity College, Dublin}
\author{Chris J. Nelson}
\affiliation{European Space Agency (ESA), European Space Research and Technology Centre (ESTEC), Keplerlaan 1, 2201 AZ, Noordwijk, The Netherlands}

\author{Gerry Doyle}
\affiliation{Armagh Observatory and Planetarium, College Hill, Armagh BT61 9DG, N. Ireland}

\begin{abstract}
Solar flare ribbons are extensive brightenings in the chromosphere during flares, often showing fine scale structuring that reflects the underlying energy release. Using high cadence imaging from the Swedish 1-m Solar Telescope/CRISP during an X1.5-class limb flare on 10 June 2014, we identify and track 232 coherent, thread-like substructures, which we term for the first time ``riblets''. From a statistical analysis, riblets have well defined lifetimes and plane-of-sky speeds (typically 5--15~s and 50--150~km/s respectively), establishing them as distinct ribbon substructures. From analysis of their temporal distributions, their distance--time (X--T) evolution uniquely reveal approximately linear and non-linear (accelerating/decelerating) classes, a discrepancy that may be influenced by projection geometry. From analysis of their spatial distributions, we find no clear correspondence between the properties of adjacent riblets, suggesting that local atmospheric conditions (fine-scale thermodynamic and/or magnetic structuring) govern their kinematics more than spatial variations in electron-beam energy flux. From analysis of their spectral distributions, clusters of riblets do show temporal and spatial coincidence with hard X-ray emission signatures, consistent with episodic electron-beam injection into the chromosphere. Using Fermi/GBM spectroscopy, we derive thick-target parameters suitable for flare simulations, with representative values $\delta_{\rm low}\approx 5.93$, $E_{\rm c}\approx 24.7$~keV, and an implied beam energy flux $\mathcal{F}_{\rm beam}\approx 1.5\times 10^{10}$~erg~cm$^{-2}$~s$^{-1}$ (based on RHESSI footpoint area). Together, these results identify riblets as the fundamental building block of flare ribbons and provide quantitative constraints for forward tests of riblet formation mechanisms.
\end{abstract}

 
\newpage
\section{Introduction} \label{sec:intro}

Carrington and Hodgson reported the first observation of a solar flare ribbon on 1st September, 1859 \citep{carrington_description_1859}. Ribbons are intense brightenings in chromospheric spectral lines and continua, produced by the sudden release of energy during solar flares in the form of electromagnetic radiation, accelerated particles, and plasma ejections.
 Ribbons are believed to form at the lower atmospheric footpoints of coronal loops, and to be a result of the flare reconnection and energization process. The brightening and expansion of these ribbons have been observed to correspond well with impulsive hard X-ray (HXR) bursts \citep{fletcher_observational_2011,benz_flare_2017, brown_solar_1980}.
Typically, within 1 second of the peak of the HXR emission, the chromosphere experiences a pronounced upward motion (chromospheric evaporation), coinciding with a rapid increase in temperature and density within the flare loop arcade \citep{fisher1985b, falchi_chromospheric_2002, sellers2022}. This initial upward motion marks a potential chromospheric signature of flare energy deposition, followed by persistent downflows in the low chromosphere driven by strong temperature gradients during the cooling phase (chromospheric condensation). \citep{fisher_dynamics_1989,abbett_dynamic_1999}.

The observation of strong brightenings in the Hydrogen-alpha 6562.8 \AA \ (herein referred to as H$\alpha$) spectral line has been a significant aspect of flare studies since their early days.
\citet{dodson_flares_1956} were among the earliest to report differences in emission profiles within flare ribbons, indicating sub-structures within the ribbons themselves. \citet{harvey_flare_kernels_1971} also observed sub-structures within flare ribbons, where small impulsive kernels exhibited distinct time-intensity profiles, compared to other parts of the ribbon. Recent numerical studies have rekindled interest in investigating the small-scale features of these kernels, revealing strong downflows \citep{druett_beam_2017} and plasma characteristics adjacent to flare kernels \citep{osborne_flare_kernels_2022}. While red-wing enhancements are often associated with emission from down-flowing material in H$\alpha$, caution must be exercised in making such assumptions, such as the effect of opacity on the interpretation of the spectral line asymmetries  \citep{kuridze_Halpha_asymmetries_2015}. However, a statistical characterisation of how widespread such fine-scale ribbon-front structures and their flow signatures are, and of the distribution of their kinematic properties during the impulsive phase, is still lacking.

 \cite{asai_evolution_2003} quantitatively evaluated the released energy during flares, emphasizing the connection between ribbon dynamics and energy release processes. \cite{Simoes2015} suggests that collisional beam-heating can only partially explain the energy demands of the 10 MK plasma observed in flare ribbons. \cite{kerr_spatial_variation_2026} used spatially resolved UV spectroscopy together with radiation-hydrodynamic modelling to show that the dominant energy transport mechanism can vary along a single flare ribbon, with some locations consistent with particle precipitation and others better explained by enhanced thermal heat flux.

High energy emission is tightly coupled to the earliest flare ribbon brightenings. Using Hinode and RHESSI (Reuven Ramaty High Energy Solar Spectroscopic Imager), \citet{Krucker2011} showed sub arcsecond co-spatiality between white light ribbon kernels and 25–100~keV hard X‐ray (HXR) footpoints, consistent with a thick target interpretation. Extending to a statistical sample, \citet{Kuhar2016} found strong co‐temporal correlations between RHESSI HXR flux (peaking for $\sim$30–50~keV photons) and excess white‐light emission, with white light kernels co‐spatial with HXR sources. At UV wavelengths, IRIS spectroscopy during SOL2014-09-10 demonstrated that chromospheric condensation and continuum enhancements track the impulsive high energy signatures in time \citep{graham_spectral_2020}. Related ribbon rooted dynamics at upper chromospheric/transition region temperatures have also been reported in \citet{Li2019}. They identified narrow, jet‐like features rooted in flare ribbons in IRIS 1330\,\AA\ images, and simultaneous Fe\,\textsc{xxi} blueshifts with Si\,\textsc{iv} redshifts indicative of chromospheric evaporation. Together, these results motivate the use of high energy contours as markers for the location and timing of energy deposition relative to ribbon formation. Nevertheless, coordinated statistical comparisons between HXR-inferred nonthermal electron-beam properties and the dynamics of resolved ribbon substructures remain scarce, the present study exploits simultaneous high-cadence imaging and high-energy context to address this gap.

In this paper, we present a novel investigation of X-class flare ribbon substructure using high cadence, wide band imaging from CRisp Imaging Spectro-Polarimeter \citep[CRISP]{scharmer_crisp_2008,lofdahl_sstred:_2021} at the Swedish 1-meter Solar Telescope \citep[SST]{sst}. Our observations uniquely capture the flare rise phase at the limb, providing a rare side on view of the chromospheric footpoints. This perspective reveals inward moving intensity thread like features (riblets) that had previously only been inferred from Dopplergrams, enabling a statistical study of their dynamics and relation to the overall ribbon evolution.

 In Section 2 we describe the observational data sets and preprocessing, the riblet identification and kinematic tracking methodology. Section 3 presents the two types of riblet evolutions, linear and non-linear through a representative example of each. Sections 4 presents the kinematic and statistical properties of both classes and examines their spatial, temporal, and multi-wavelength context (UV/EUV and high-energy X-ray contours) and evaluates possible interdependence. Section 5 discusses projection effects and loop geometry implications on the observed evolution. Sections 6 and 7 discuss the results and summarise the main conclusions.

\section{Observations and Methods}\label{sec:obs}

\begin{figure*}
  \centering
  \subfloat[]{%
    \includegraphics[width=0.49\linewidth]{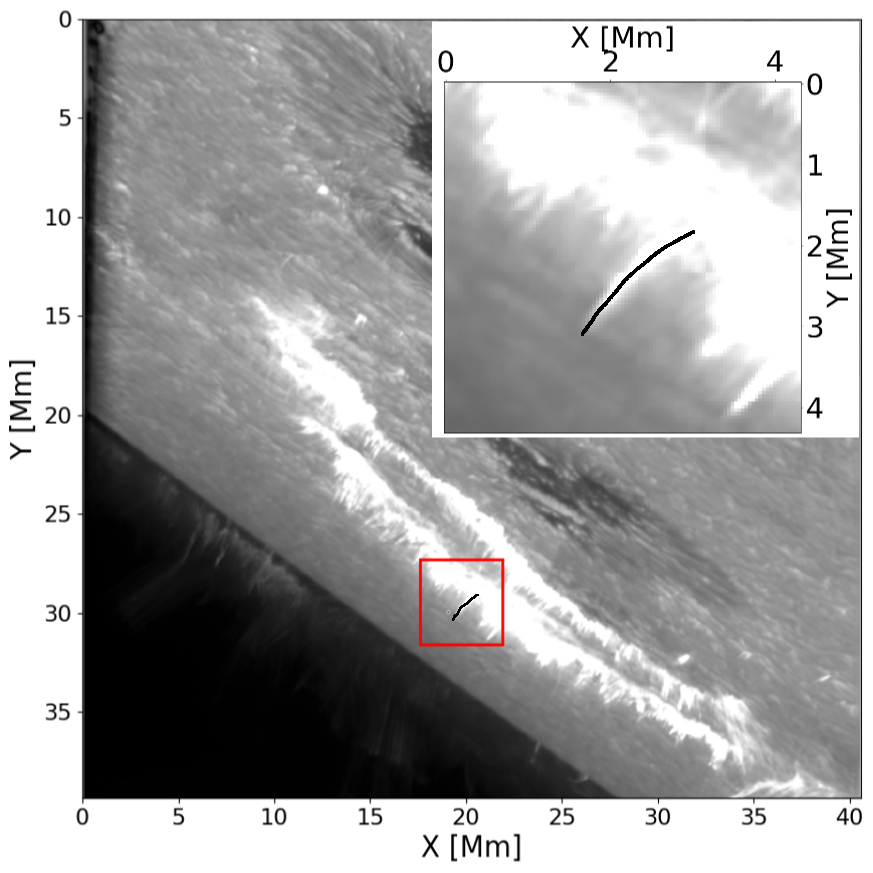}%
    \label{fig:snapshot}%
  }\quad
  \subfloat[]{%
    \includegraphics[width=0.49\linewidth]{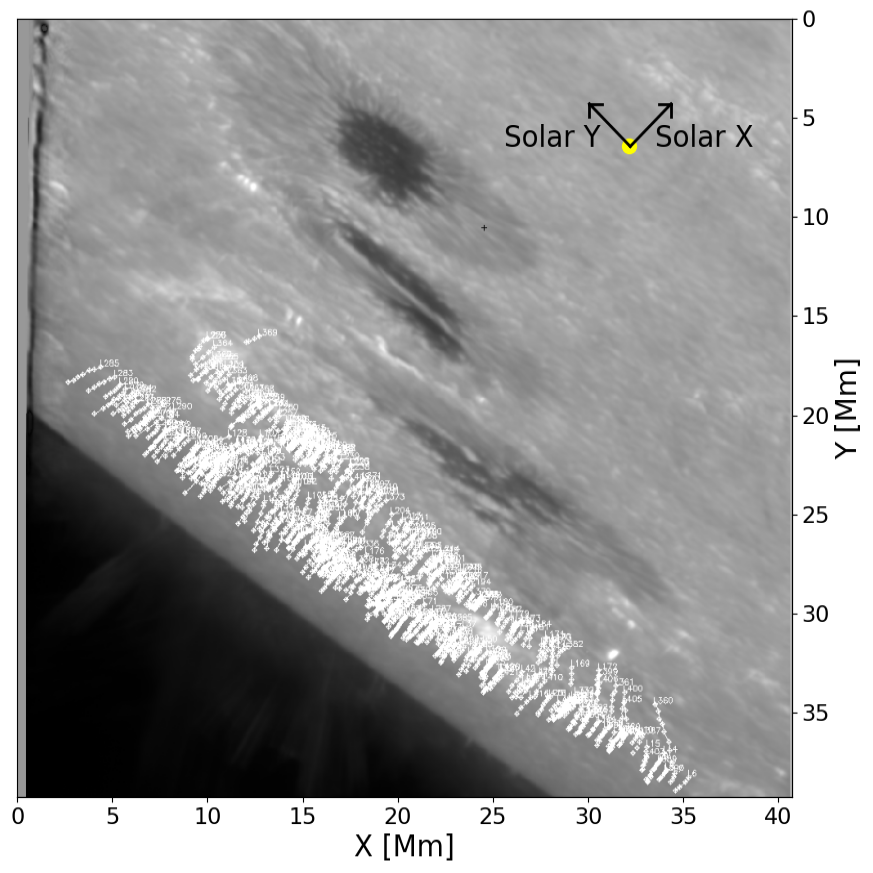}%
    \label{fig:track}%
  }

  \vskip\baselineskip
  \subfloat[]{%
    \includegraphics[width=\linewidth]{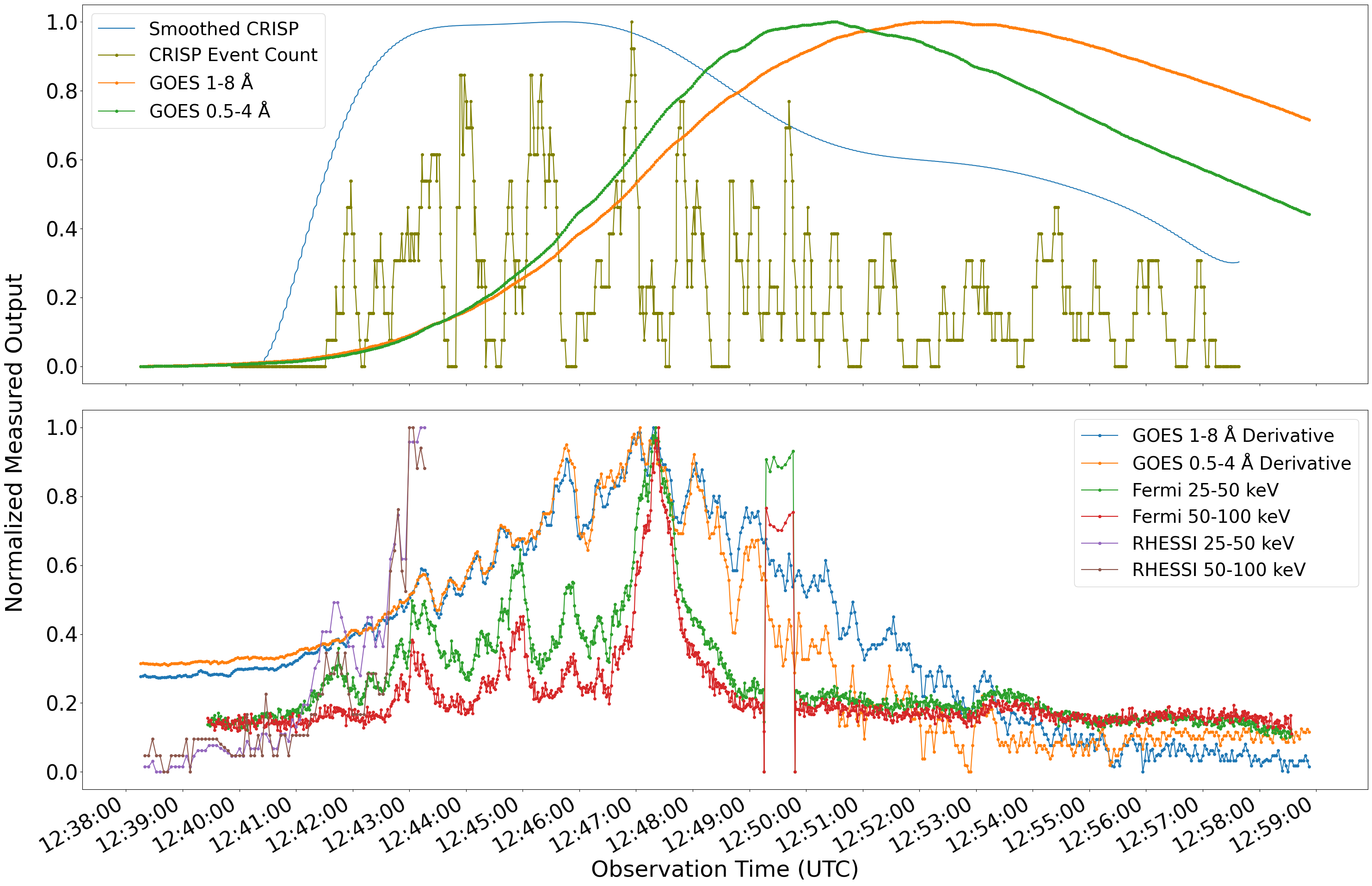}%
    \label{fig:xray_lc_all}%
  }

  \caption{Flaring event (NOAA 12087): 
   \textit{(a)} H$\alpha$ wideband SST/CRISP image of the flare (Time-12:51:53 UTC) with a red boxed region (expanded in the inset image) of a single example riblet, with track overlaid as black line. 
   \textit{(b)} 232 tracked riblets (white dotted lines). Yellow dot in the top right (for reference) has coordinates X,Y =(-857,-297) arcsec. The length of each arrow corresponds to 5 arcsec. 
   \textit{(c)} Lightcurves of CRISP H$\alpha$ (smoothed), Fermi, RHESSI, GOES, GOES derivatives and event count of riblets.
  }
  \label{fig:images}
\end{figure*}

With high spatial and temporal resolution observations of the impulsive phase of a limb X-class flare, the ribbons appear as small scale, short lived, thread like substructures along their entire length. These substructures manifest as plasma columns, characterized by a gradual decrease in intensity as they shrink, grow, or traverse in directions out of the plane of the main flare ribbon body before eventually merging back into the main body. We shall henceforth refer to these phenomena as \textit{riblets}.

\subsection{X1.5-class Flare Overview}
On 10 June 2014 at 12:52:00 UTC, an X1.5-class flare occurred at the east limb (S20E89) in NOAA AR 12087. We observed the event with SST/CRISP in H$\alpha$ at an image scale of 43 km pixel$^{-1}$.
CRISP at the SST is a dual-etalon Fabry-Perot imaging spectropolarimeter with a 55$\times$55 arcsec field of view (FOV). The wideband CRISP camera has a passband of 6\AA.
During the flare, the correlation tracker and adaptive optics were intentionally disabled. Enhanced ribbon brightenings near the limb caused erratic pointing corrections, switching off the systems yielded more stable tracking and continuous monitoring of ribbon evolution under relatively good seeing.
The observing cadence varied from 0.2 s to 1.9 s. Cadence fluctuations arose from the disabled tracker and from exposure adjustments by the telescope chopper in response to large photometric changes. We account for this by using the per-frame timestamps in all analyses.
Data were reduced and post-processed with the CRISPRED pipeline \citep{de_la_cruz_rodriguez_crispred:_2015}, employing MOMFBD \citep{mats02,2005SoPh..228..191V} and a corrective technique for differential stretching. SST data were coaligned to the SDO (
Solar Dynamics Observatory) heliocentric coordinate system to enable spatio-temporal comparisons with AIA (Atmospheric Imaging Assembly ) and RHESSI.
Figure \ref{fig:snapshot} shows a CRISP snapshot with an inset highlighting a single riblet, the scale bar uses $1\arcsec \approx 0.725$ Mm. Figure \ref{fig:track} overlays the tracks of all tracked riblets across the full H$\alpha$ FOV and throughout the entire observing sequence. To place the event in broader context, Figure \ref{fig:images}(c) presents intensity light curves from CRISP together with high-energy instruments co-temporal measurements. The smoothed CRISP H$\alpha$ wideband curve is computed, for each frame, as the mean of pixels with intensity greater than the mean riblet intensity defined in Figure \ref{fig:mask3}, sudden dips correspond to frames affected by poor seeing. Effectively, this means that the curve represents the intensity averaged riblet curve that will include intensities beyond the pixels that are used to represent the riblet tracks
We also use Fermi/GBM (Gamma-ray Burst Monitor) \citep{meegan_fermi_2009} and RHESSI \citep{lin_reuven_2002} to compare high-energy emission with the chromospheric signal in the riblets. The CRISP and Fermi/GBM hard X-ray peak precedes the GOES soft X-ray peak, consistent with the Neupert effect \citep{Neupert1968}. Here we can confirm that, since the riblet signature rises in accordance with the HXR emission we assume that the riblet formation shares a driver with the impulsive HXR emission.

\subsection{Riblet Detection and Image Processing} \label{track}
With the exceptional spatial and temporal resolution of the data at our disposal, we can analyze the motion and behaviour of riblets in unprecedented detail. However, the nature of the observation introduces a significant challenge in developing an automated approach for detecting and tracking riblets. The acquired images are frequently characterized by considerable seeing-induced noise, and the riblets themselves tend to manifest in close proximity to one another, and sometimes obscured by the saturated ribbon pixels, appearing more like a ribbon blob than as a discrete thread-like feature. In view of this, we performed a manual tracking of the evolving riblets throughout the duration of the observation where only thread like structures were selected and any blobs showing similar motion were not considered for this study. We utilized CRISPEX \citep{vissers_flocculent_2012}, an interactive data language (IDL) based graphical user interface, specifically designed for the interactive and real-time analysis of SST observations, to trace all thread-like protrusions originating from the ribbon. Specifically, we selected instances where the complete evolution of these protrusions, including their retraction back into the ribbon base, could be observed, while excluding cases involving frames exhibiting severely compromised seeing quality. As a result of our analysis, we were able to successfully identify a total of 232 events. This analysis was performed on the H$\alpha$ wide-band images of the flare because they show the photosphere in quiet regions and the riblets in the flare. This enabled us to more confidently capture measurements of riblet lengths, without introducing obstructions caused by other chromospheric features such as fibrils or fan jets.

Before proceeding with riblet intensity computation and masking, we conducted an estimation of the seeing quality for each frame in our data set. To accomplish this, we selected a region in the observation with little to no activity throughout the entire duration (an area of 500x500 pixels). The seeing quality of each frame is represented by the ratio of the standard deviation to the mean of intensity values within this designated area (higher value means better seeing). By employing this approach, we are able to identify and retain image frames with high-seeing quality while filtering out those with low seeing quality.

\begin{figure*}
  \centering

  \subfloat[]{
    \includegraphics[width=0.45\textwidth]{ 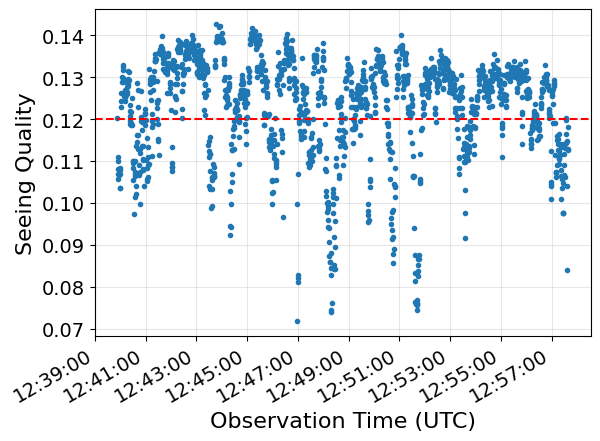}
    \label{fig:scene}
  }\quad
  \subfloat[]{
    \includegraphics[width=0.45\textwidth]{ 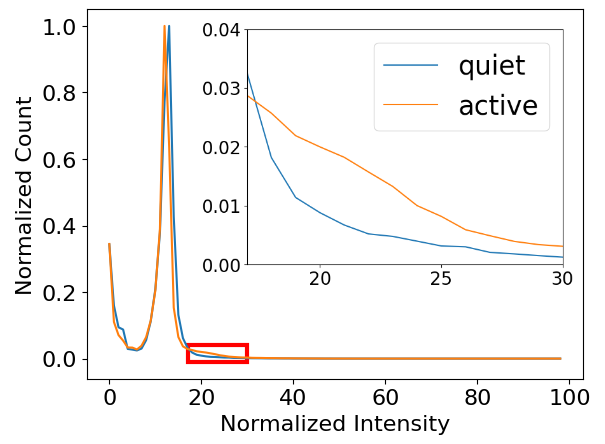}
    \label{fig:mask1}
  }

  \vskip\baselineskip

  \subfloat[]{
    \includegraphics[width=0.45\textwidth]{ 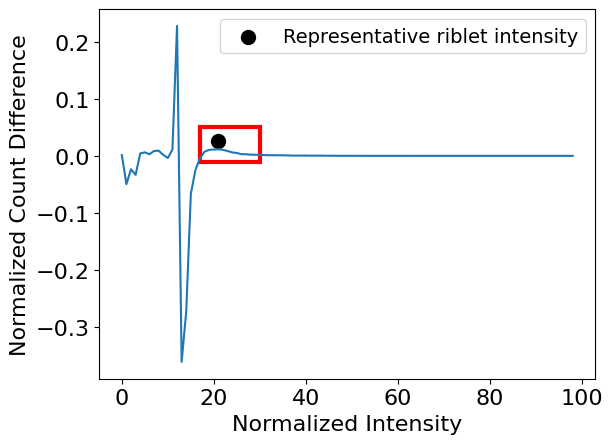}
    \label{fig:mask2}
  }\quad
  \subfloat[]{
    \includegraphics[width=0.45\textwidth]{ 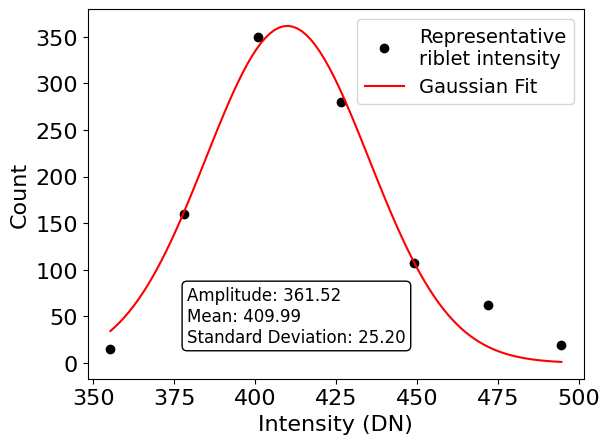}
    \label{fig:mask3}
  }

  \caption{Steps to calculate masking intensity for the riblets. 
   \textit{(a)} Seeing quality of the frames of the observation on 10 June 2014. The red line marks the threshold (0.12) for frames used for the riblet intensity calculation. 
   \textit{(b)} Intensity histogram of a quiet frame with no riblets (blue) and an active frame with riblet activity (orange). Expanded view of the area in the red box to show the difference in intensity values that represent the riblets in the frame. 
   \textit{(c)} Difference between the quiet and active frame histograms in (b). Peak count intensity was chosen as representative riblet intensity for the frame. The red box is the same as in (b). 
   \textit{(d)} Gaussian fit of the peak count value of riblet intensities for all active frames with seeing quality above the threshold.
  }
  \label{fig:masking}
\end{figure*}

The seeing quality of all frames is illustrated in Figure \ref{fig:scene}. In our computations, we have taken into account frames with a seeing quality of 0.12 or higher, thereby encompassing 69.9\% of the total frames. This threshold was chosen as it provided a substantial number of usable frames while ensuring acceptable visual quality upon inspection. Additionally, with this cutoff, the largest time interval between successive frames of acceptable quality is 41 frames, corresponding to an maximum gap of approximately 8.2 seconds in the most severely affected period of the sequence. The nominal observing cadence varied between 0.25 s and 1.9 s, and even after filtering for seeing, the median time interval between frames remained significantly shorter than the riblet lifetimes. These selected frames with adequate seeing quality will be employed in the subsequent steps of our analysis, facilitating a robust investigation of riblet intensities and their associated dynamics.

After manually tracking the riblets, we can construct distance-time plots (X-T cuts) to visualize the temporal evolution of the pixel intensity along the riblet tracks pixel-by-pixel. These X-T cuts allow us to analyze kinematics of the riblets in time as well as establish whether or not there are repeat events along the same trajectory, potentially giving us further insights into the behaviour of riblets. However, to obtain accurate riblet intensities, it is necessary to separate the riblets from the background in the X-T cuts. This can be accomplished by determining the range of intensity values for the riblets and using those values to generate a threshold for what constitutes time averaged riblet intensities and then to establish a mask.

To determine the ribbon intensity, we adopted the following procedure (presented in Figure \ref{fig:masking}). We selected an image with minimal ribbon activity (pre-flare) and another image with significant ribbon activity and plot their respective intensity histograms (Figure \ref{fig:mask1}). Because these histograms are normalized to the total pixel count within the field of view, the emergence of the bright flare ribbon population (at higher normalized intensities) causes a corresponding decrease in the normalized count of the background population compared to the quiet-Sun state. This does not imply the background itself has darkened, but rather that it occupies a smaller relative proportion of the active-frame image area. In selection of the background pixels, we chose a region that was far away from the ribbon formation but at a similar proximity to the solar limb, so that the background intensity could accurately represent the background of the flaring atmosphere. We then computed the difference between the two histograms and isolated the residual (a peak in the distribution of intensities above the mean), representing the ribbon activity, which is highlighted in red in Figure \ref{fig:mask2}. Subsequently, we determined the intensity value corresponding to the highest count from the isolated peak. We repeated this process for all frames with both adequate seeing quality ($>$0.12) and the presence of ribbon activity, utilizing the same reference frame with minimal ribbon activity.

Finally, we fitted the histogram of calculated intensity values with a Gaussian profile of the form:

\begin{equation}
 y= a\cdot \exp \left({- \frac {(x-\mu)^2} {2\sigma^2}}\right)
\end{equation}
where $a$, $\mu$, and $\sigma$ represent the amplitude, mean and standard deviation of the Gaussian profile, respectively (Figure \ref{fig:mask3}). We noted that a large number of these frames resulted in the same intensity values, and although we returned a gaussian distribution as expected, we only had seven unique intensity values for the fit.

This process yielded a mean ribbon intensity of $\mu=412.60$ and a standard deviation of $\sigma=24.05$. We adopted these values to mask out the riblets in the X-T cut data, from which we could then distinguish between background and riblet activity. This fitting process allowed us to accurately estimate the ribbon intensity for each frame and generate a mask for the riblets in the X-T cuts with a  region of uncertainty between the riblets and the background (Figure \ref{fig:linear}b and \ref{fig:nonlinear}b). The spatial extent of this uncertainty region (defined by the $\mu\pm\sigma$ intensity range) provides the error bars for our subsequent kinematic tracking, accounting for the intensity gradients at the riblet tip. The resulting masks are then used for further analysis of the riblet kinematics. The reliability and stability of the masking values were assessed through a parameter survey study. Specifically, the parameters $\mu$ and $\sigma$ were systematically modified within the ranges of $\mu+\sigma$ to $\mu-\sigma$ and $2\sigma$ to $\sigma/2$, respectively. This approach allowed us to examine the impact of different parameter combinations on the resulting riblet velocities that are derived from the masked regions of the X-T cuts. The outcome of this procedure demonstrated a high degree of robustness in the masking process. We found that the velocities obtained were not affected significantly by the choice of these parameters $\mu$ and $\sigma$. In the next section, we use these masked X-T cuts to quantify riblet boundary kinematics and to define the classification scheme that separates approximately constant-speed (linear) events from accelerating/decelerating (non-linear) events.

\section{Riblet Classifications}
\begin{figure*}
     \centering
         \centering
         \includegraphics[width=\textwidth]{ 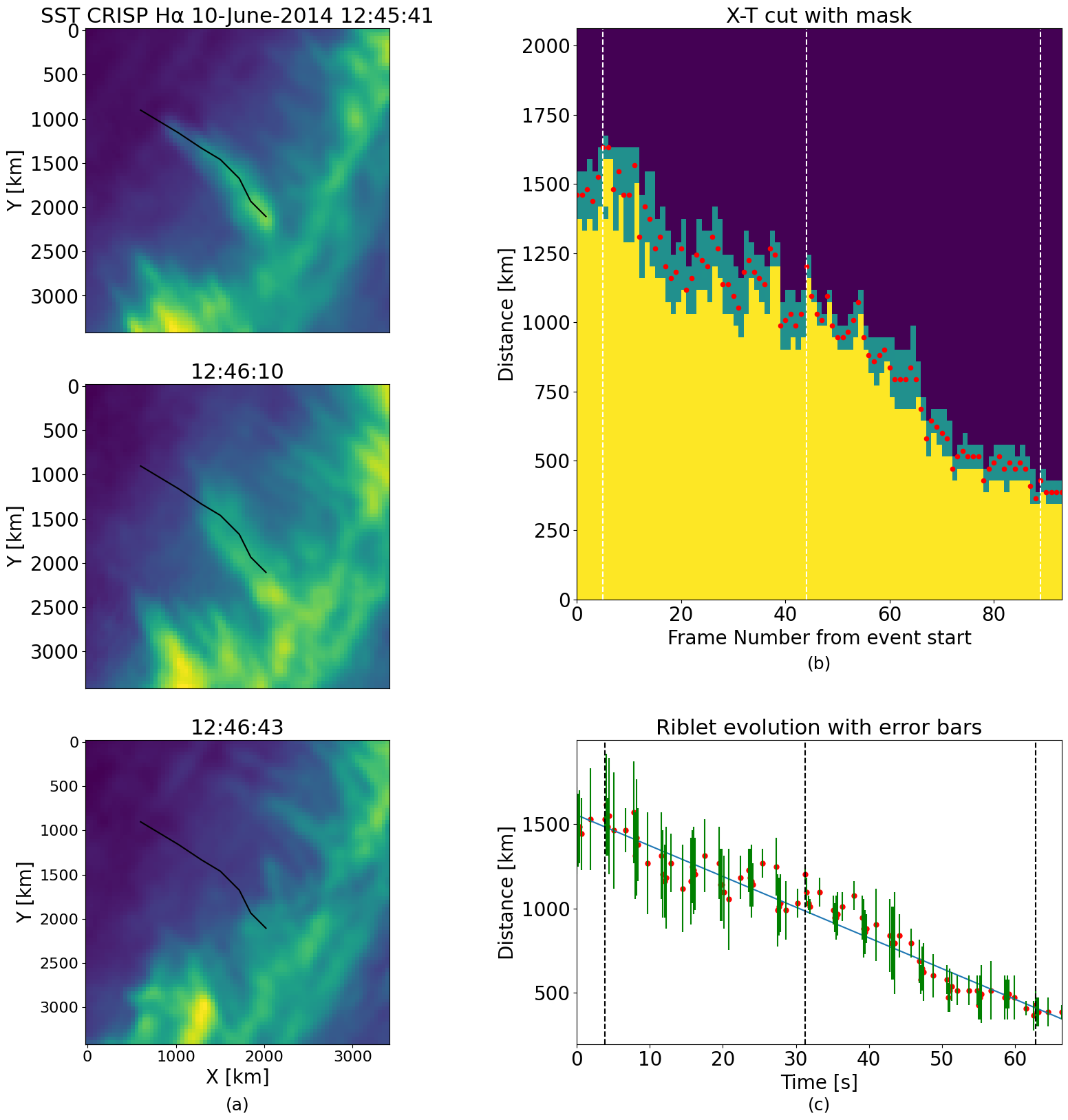}
         \caption{Riblet evolution of a typical constant speed linear riblet.\textit{ (a)} Frames with the tracked riblet (black line) at different times in the life of the riblet.\textit{(b)} Evolution of intensities of the tracked pixels (riblet) over time. The red dots represent the tip of the riblet, as determined by the masking technique. The green section indicates the region of uncertainty between the riblets and the background during our observation ($\mu\pm\sigma$). The yellow section represents the signal ($>\mu+\sigma$), while the violet section represents the background ($<\mu-\sigma$). The vertical dashed white lines indicate the time stamps of the frames in (a).\textit{(c)} The evolution of the length of the riblet with time. Tip of the riblet (red dots), uncertainty in the length (green bars) and a first-order polynomial fit (blue line). The vertical black dotted lines point out the time of the frames in (a). }
         \label{fig:linear}
     \end{figure*}

We observed two distinct classifications of riblets within the kinematic analysis using the X-T cuts, with one class exhibiting constant speed profiles (linear riblets)  and the other showing either acceleration or deceleration profiles (non-linear riblets). Figure \ref{fig:linear}a provides a visual representation of a linear riblet, consisting of three snapshots taken at the beginning, middle, and end stages (top to bottom panels). Figure \ref{fig:linear}b displays the masked X-T cut data, which portrays the intensity evolution of the tracked pixels extracted from the pixels along the black lines in Fig \ref{fig:linear}a. To isolate the signal from the background, we apply the intensity threshold values obtained in Section \ref{track}. The red dots represent the riblet leading edge boundary, as determined by the masking technique. The spatial extent of the green section is a measure of the uncertainty of the location of the boundary, i.e. representing the intensity transition between the riblet and the background in our observation. The yellow band pixels represent the riblet signal above the background intensity, whilst the violet band pixels represent the background. Figure \ref{fig:linear}c shows only the riblet's boundary profile, including a fitted polynomial function and associated fit error bars, excluding the mask. This plot also takes into account the irregular time intervals between frames of the observation overlaid on a continuous time axis.
 Every riblet boundary profile was fitted with a second-order polynomial using the \texttt{polyfit} function from the \texttt{NumPy} Python module. From this fit, we obtained the acceleration, along with its associated error. Linear riblets are defined here as those for which the absolute value of acceleration minus its error was consistently zero, whereas non-linear riblets are defined as those which exhibited statistically significant acceleration magnitudes including the error margin above zero.

We fit each riblet with a quadratic function of the form:
\begin{equation}
x(t) = c + v\, t + \frac{1}{2} a\, t^2 ,
\end{equation}
where \(x(t)\) is the riblet length as a function of time \(t\), \(c\) is the initial length at \(t = 0\), \(v\) is the initial speed, and \(a\) is the acceleration. From this fit, we obtain the acceleration coefficient \(a\) along with its associated uncertainty \(\sigma_a\).
 We define the z-statistic as
\begin{equation}
Z = \frac{|{a}|}{\sigma_{{a}}}.
\end{equation}
We classify
\begin{eqnarray}
& &\text{linear} \leftrightarrow Z < 1 \;\; \big(|a| \le \sigma_{a}\big),
\nonumber \\
& &\text{non\text{-}linear} \leftrightarrow Z \ge 1 .
\end{eqnarray}
Thus, a riblet may have a nonzero $a$ numerically yet be classified as linear if its acceleration is statistically indistinguishable from zero within $1\sigma$.

 Next, when we establish a linear riblet we fit the riblet boundary with a first-order polynomial to determine the speed (Figure \ref{fig:linear}c). In the example in Figure \ref{fig:linear}, the speed determined by a first-order polynomial fit is 18.04 km/s.

\begin{figure*}
         \centering
         \includegraphics[width=\textwidth]{ 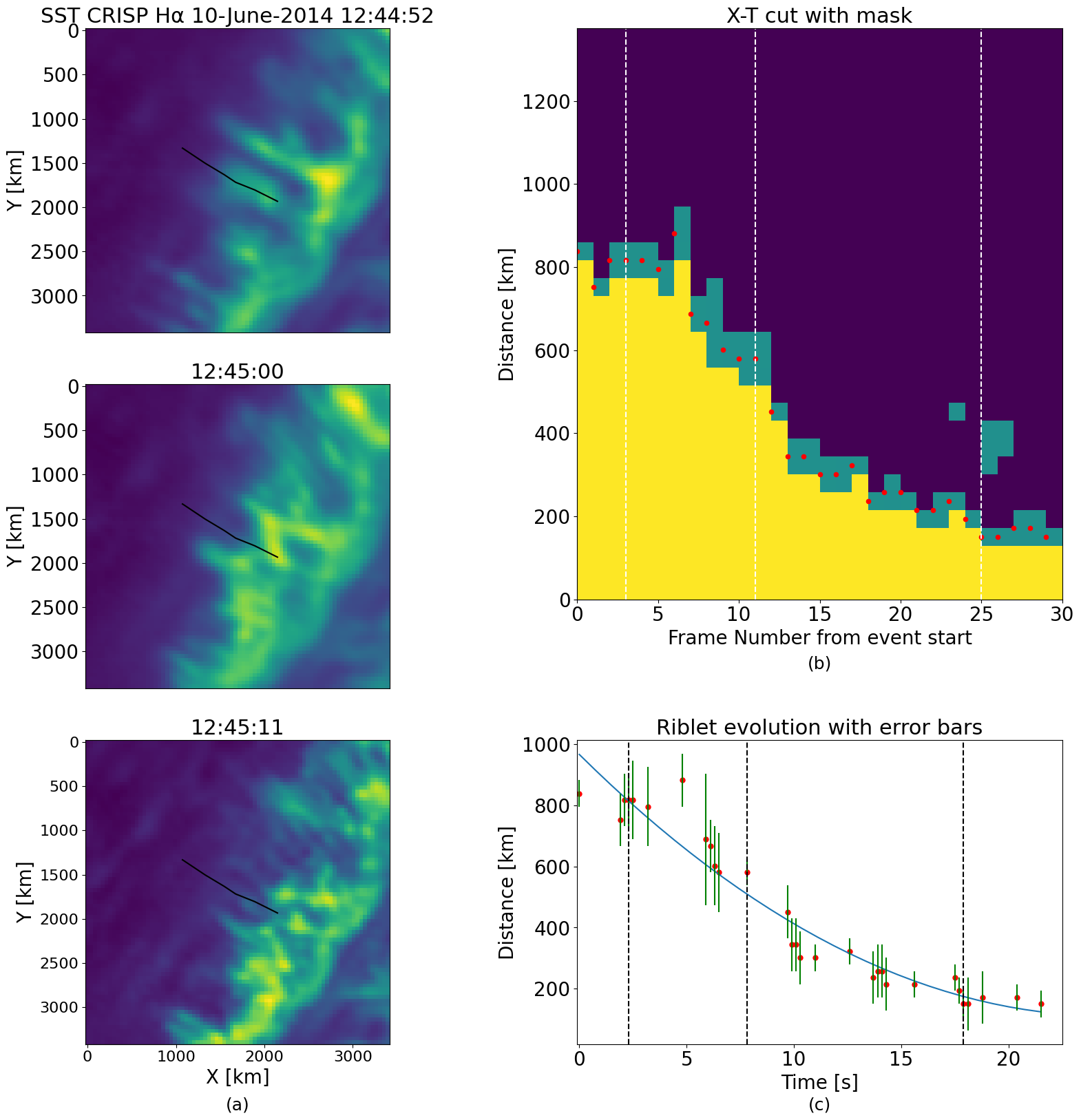}
         \caption{Riblet evolution of a typical accelerated non-linear riblet.\textit{(a)} Frames with the tracked riblet (black line) at different times in the life of the riblet.\textit{(b)} Evolution of intensities of the tracked pixels (riblet) over time. The red dots represent the tip of the riblet, as determined by the masking technique. The green section indicates the region of uncertainty between the riblets and the background during our observation ($\mu\pm\sigma$). The yellow section represents the signal ($>\mu+\sigma$), while the violet section represents the background ($<\mu-\sigma$). The vertical dashed white lines indicate the time stamp of the frames in (a).\textit{(c)} The evolution of the length of the riblet with time. Tip of the riblet (red dots), uncertainty in the length (green bars) and a second-order polynomial fit (blue line). The vertical black dotted lines point out the time of the frames in (a).}
         \label{fig:nonlinear}
\end{figure*}

In the non-linear riblet example  (Figure \ref{fig:nonlinear}),the fit provides us with an initial speed of 69.43 km/s and acceleration of -0.70 km/s$^{2}$($Z=\frac{|a|}{\sigma_a}=\frac{0.70}{0.36}=1.94>1$, hence non-linear). To get an estimate of the average speed of the riblet during the course of the evolution, we also implement a linear fit, which provides us with an average speed of 39.41 km/s. This linear fitting approach was applied to the non-linear example in order to draw a fair comparison between the linear and non-linear examples.

Table \ref{table1} presents the fit parameters for representative linear and non-linear riblet events. This is the first time that the sub-structures of the ribbons have been classified in this way. It may be indicative of either different driving mechanisms or different local atmospheric conditions giving rise to different riblet formation responses, or potentially geometric viewing angle effects arising from non-radial projects of the riblets. Given that we can apply this analysis to many riblet events we can begin to better understand the underlying physical basis for the classifications from a statistical standpoint.

\begin{table}[]
\centering
\begin{tabular}{|l|l|l|}
\hline
                                                & {Linear}     & {Non-Linear} \\ \hline
Observed                              & 67.5                & 22.8                \\
Lifetime (s)                          &                     &                     \\ \hline
Observed Initial                               & 1462.0              & 838.5               \\
Length (km)                            &                     &                     \\ \hline
Fit Initial                            & 1496.46 $\pm$ 23.83 & 965.8 $\pm$ 36.89   \\
Length (km)                            &                     &                     \\ \hline
Fit Initial                            & -18.04 $\pm$ 1.70   & -69.43 $\pm$ 7.99   \\
Speed(km/s)                        &                     &                     \\ \hline
Fit Average                            & -0.07$\pm$ 0.02     & 0.7 $\pm$ 0.36      \\
Acceleration(km/s$^{2}$) &                     &                     \\ \hline
\end{tabular}
\caption {Parameters of 2nd-order polynomial fit of the speed profiles of the riblets shown in Figures \ref{fig:linear} and \ref{fig:nonlinear}.}
\label{table1}
\end{table}

\section{Statistical Properties of Riblets} \label{sec:stats}

In this section we quantify the statistical behavior of all the riblets and compare the linear and non-linear populations. We first summarize the distributions of the key fitted and measured parameters ( acceleration, initial length, lifetime, onset time, and characteristic speeds) using histograms to establish typical values and to assess whether the two classes are separable in parameter space. We then examine how these quantities vary during the flare rise phase, and place the riblet occurrence in multi-wavelength context using UV/EUV diagnostics and high-energy X-ray constraints. Finally, we map the same properties in the image plane to test for spatial organization along the ribbon and to evaluate whether local emission is predictive of riblet behavior. 

\subsection{Histograms of riblet properties}\label{sec:riblets_histograms}

\begin{figure*}
  \centering

  \subfloat[]{
    \includegraphics[width=0.45\textwidth]{ 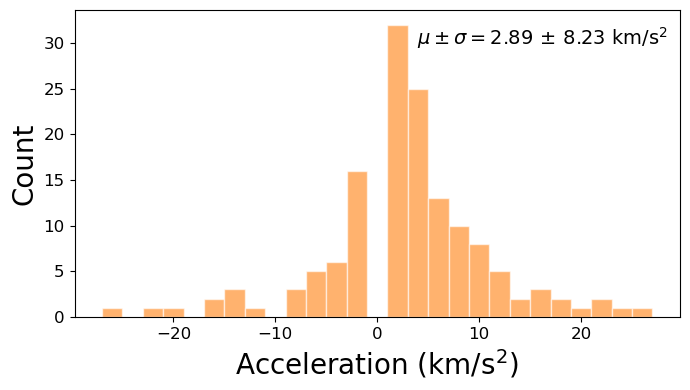}
    \label{fig:acc_hist}
  }\quad
  \subfloat[]{
    \includegraphics[width=0.45\textwidth]{ 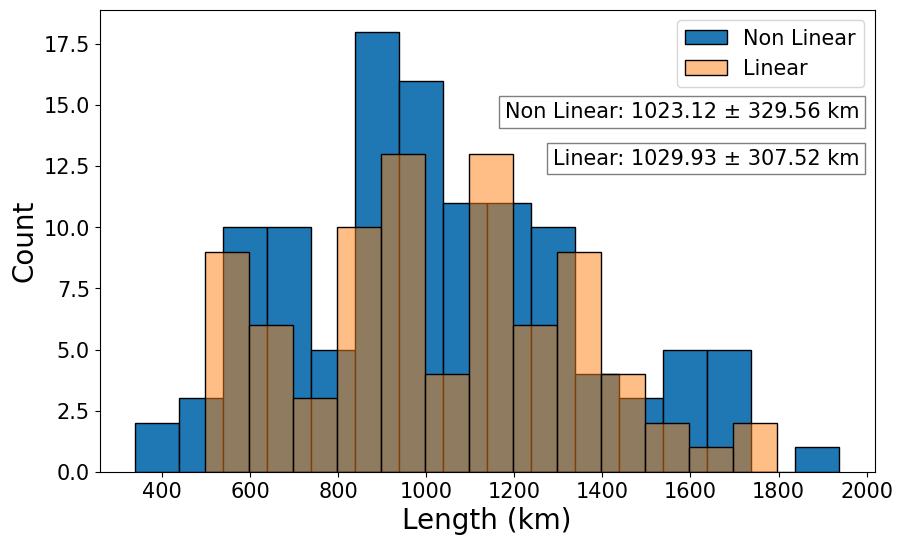}
    \label{fig:len_hist}
  }

  \vskip\baselineskip

  \subfloat[]{
    \includegraphics[width=0.45\textwidth]{ 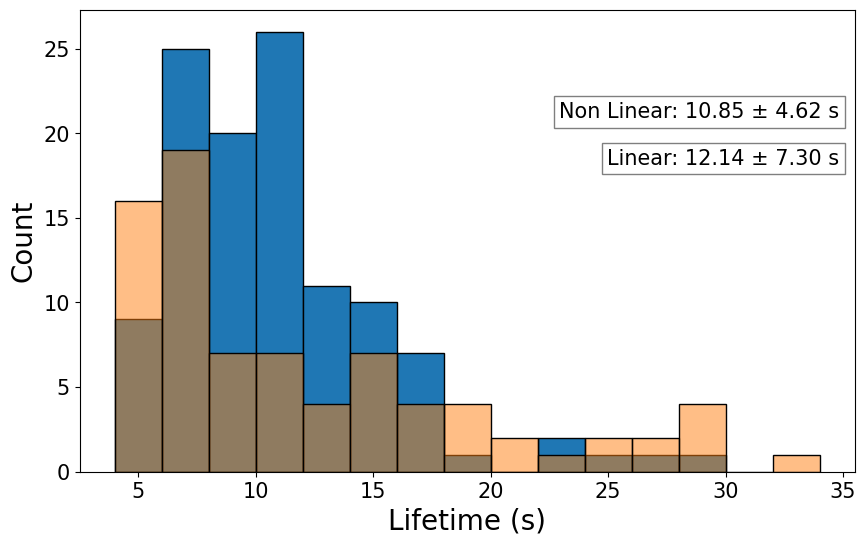}
    \label{fig:lifetime_hist}
  }\quad
  \subfloat[]{
    \includegraphics[width=0.45\textwidth]{ 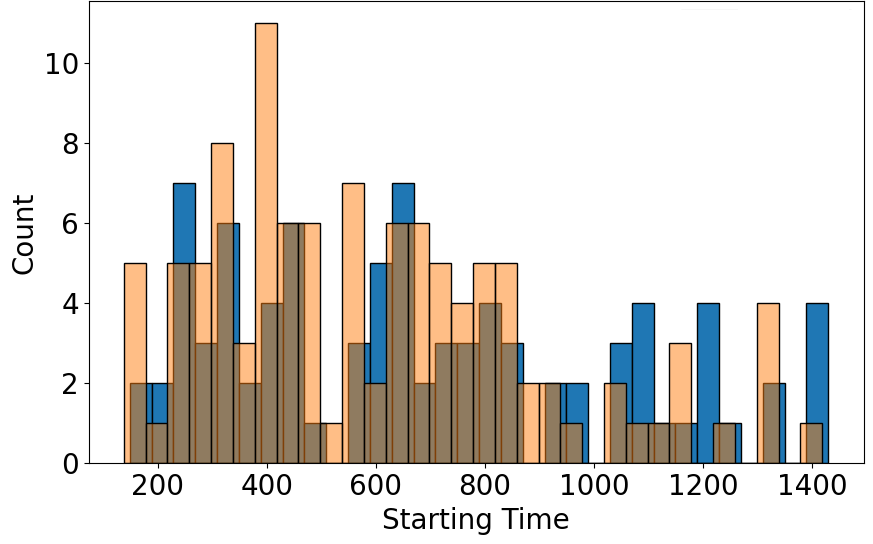}
    \label{fig:st_hist}
  }

  \vskip\baselineskip

  \subfloat[]{
    \includegraphics[width=0.45\textwidth]{ 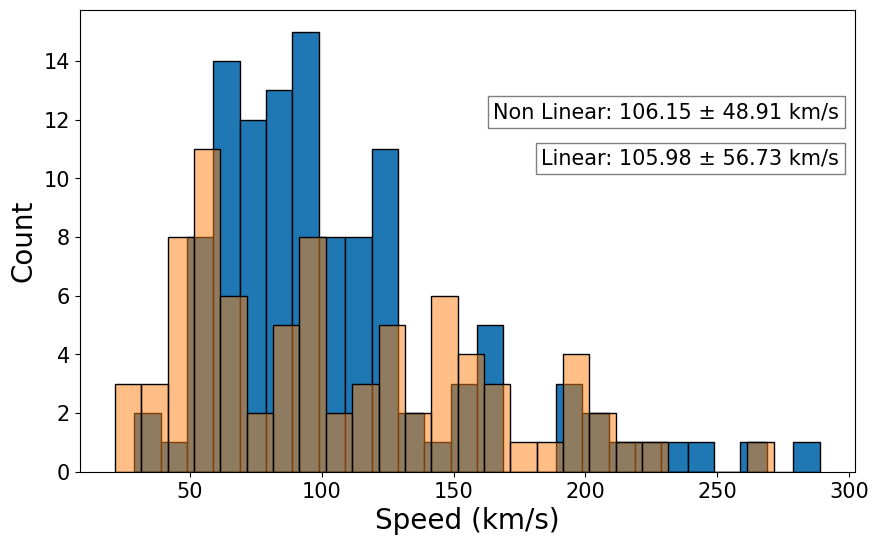}
    \label{fig:avg_speed_hist}
  }\quad
  \subfloat[]{
    \includegraphics[width=0.45\textwidth]{ 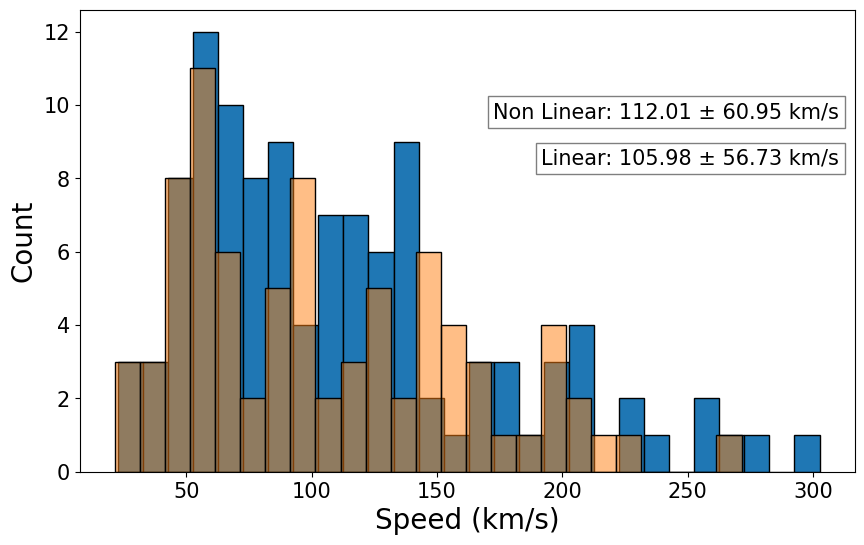}
    \label{fig:max_speed_hist}
  }

  \caption{Histograms of key riblet parameters (Non-Linear: blue, Linear: Orange) 
   \textit{(a)} acceleration values used to classify riblets
   \textit{(b)} initial lengths 
   \textit{(c)} lifetimes 
   \textit{(d)} start times with Starting time (Frame Number)  to UTC reference: 0~$\rightarrow$~12{:}39{:}52, 350~$\rightarrow$~12{:}44{:}04, 700~$\rightarrow$~12{:}48{:}18, 1050~$\rightarrow$~12{:}52{:}29, 1400~$\rightarrow$~12{:}56{:}41
   \textit{(e)} average speeds 
   \textit{(f)} maximum speeds.
  }
  \label{fig:hist}
\end{figure*}

In Figure \ref{fig:hist}, we present a comprehensive overview of the statistical properties of several key riblet parameters, enabling a comparative analysis between linear and non-linear riblet events.

Figure \ref{fig:hist} (a) shows the histogram of accelerations derived from the second-order fits.
This plot provides a quantitative overview of the dynamical nature of the riblet populations.

Figure \ref{fig:hist} (b) displays the initial lengths of the riblets, measured at the onset of their evolution. The distributions for both linear and non-linear riblets are very similar, indicating that the initial spatial extent of these features is not a distinguishing characteristic between the two classes. 

Figure \ref{fig:hist} (c) presents the lifetimes of the riblets, defined as their duration from their first appearance until they merge into the flare ribbon. Both linear and non-linear riblets exhibit similar lifetime distributions, reinforcing the observation that while their kinematic evolution profiles may differ, the lifetimes are not strongly influenced by their evolution.

Figure~\ref{fig:hist} (d) shows the histogram of riblet start times, with the x-axis denoting frame numbers corresponding to the temporal cadence of the observation. Although the overall occurrence of riblets varies during the rise phase, the temporal distributions of the linear and non-linear populations are essentially the same. This indicates that both categories of riblets are triggered under similar conditions throughout the flare evolution.

Figure \ref{fig:hist} (e) and (f) show the average and maximum speeds, respectively, of the riblets. These values were computed from the second-order polynomial fits. For consistency, we present only the second-order based results. These plots reveal that non-linear riblets tend to exhibit higher average and peak speeds than their linear counterparts enabling us to classify them on this basis. 

\subsection{Temporal properties of riblets}\label{sec:temp_dist}

\begin{figure*}
  \centering
  \subfloat[]{
    \includegraphics[width=0.45\textwidth]{ 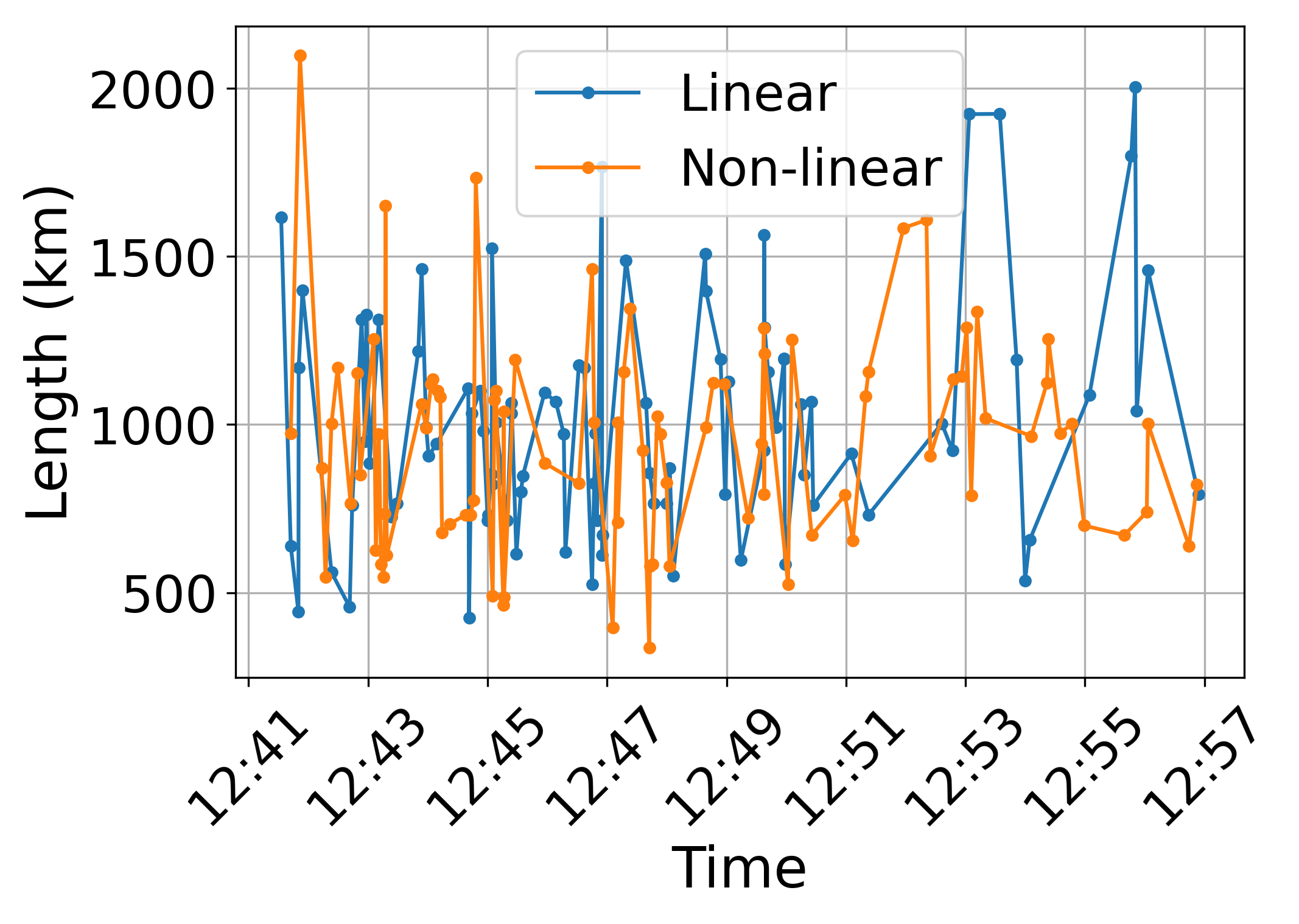}
  }\quad
  \subfloat[]{
    \includegraphics[width=0.45\textwidth]{ 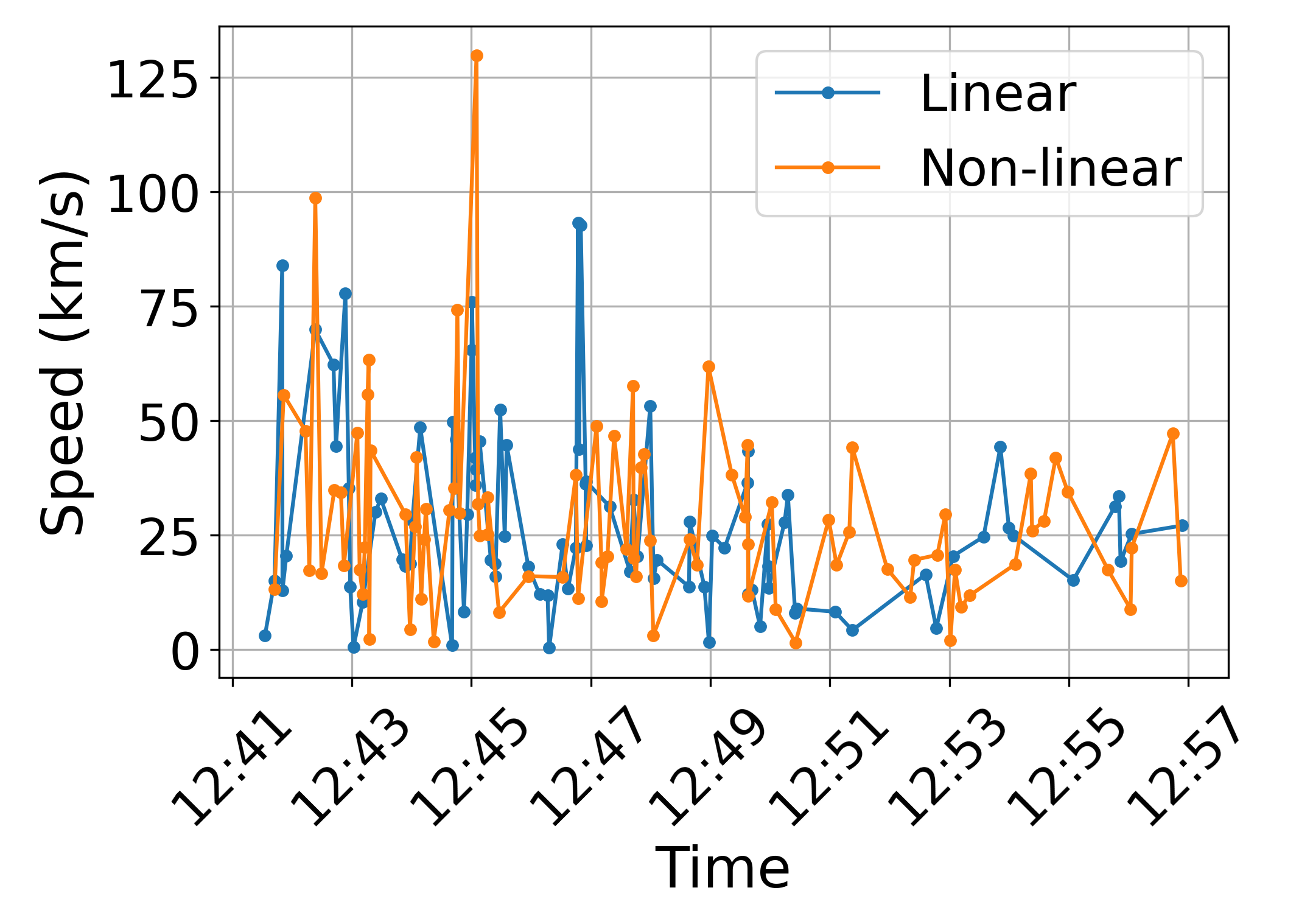}

  }

  \vskip\baselineskip
  \subfloat[]{
    \includegraphics[width=0.45\textwidth]{ 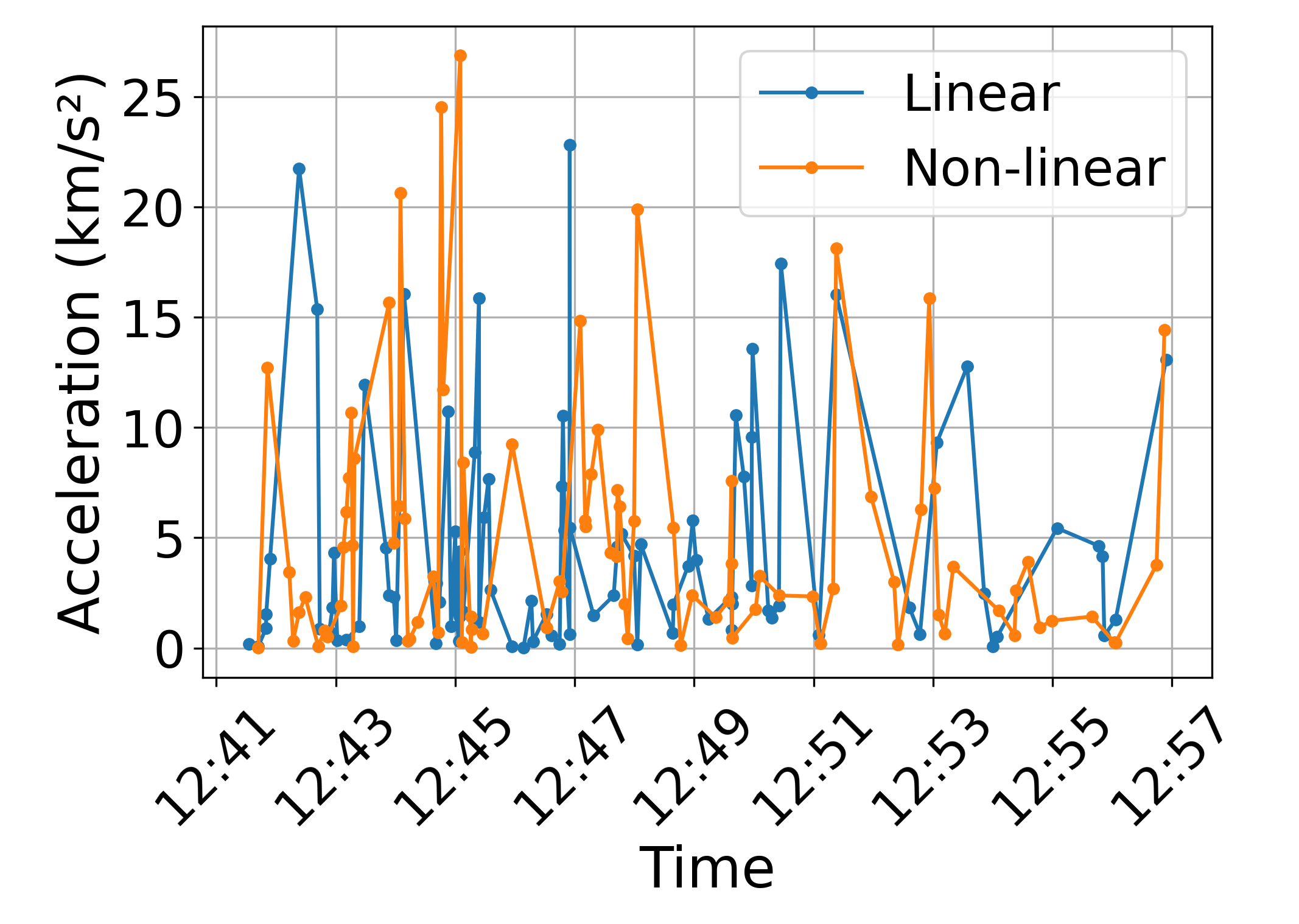}

  }\quad
  \subfloat[]{
    \includegraphics[width=0.45\textwidth]{ 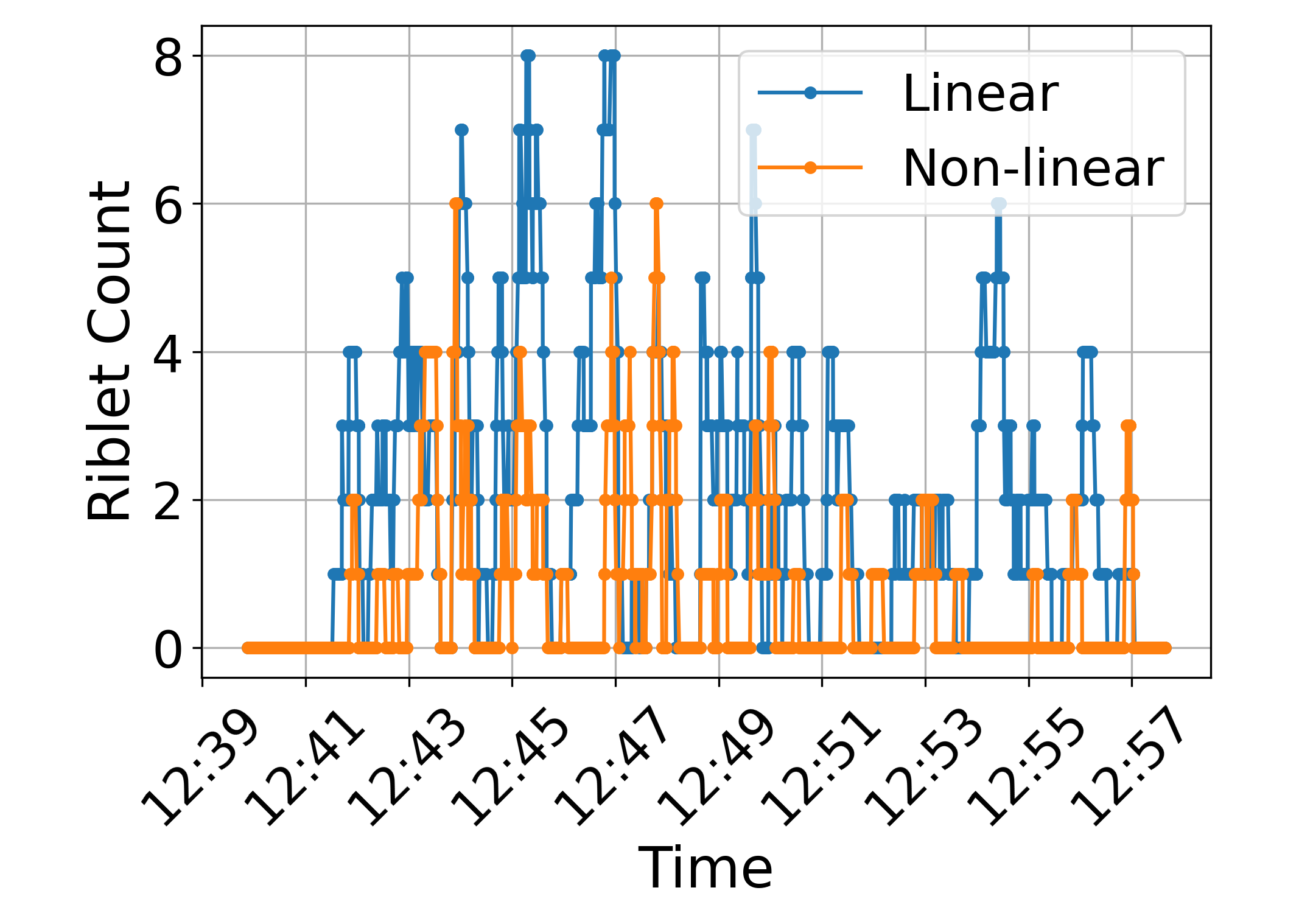}

  }

    \caption{Time series (UTC) plots of spatially averaged riblet parameters: \emph{(a)} Initial length per frame.\emph{(b)} Linear speed obtained by first-order fit. \emph{(c)} Acceleration obtained by second-order fit.\emph{(d)} Total number of riblets per frame.}
    \label{fig:ts}
\end{figure*}

Figure~\ref{fig:ts} presents a time series of spatially averaged properties for both linear and non-linear riblets over the full observation.

In figure \ref{fig:ts} (a), the average initial length of riblets is plotted as a function of time. The initial lengths of all newly emerging riblets at each time step are averaged and displayed. The absence of any discernible temporal pattern in this quantity suggests that the initial spatial extent of riblets is not systematically dependent on their time of appearance during the evolution of the flare.

Figure \ref{fig:ts} (b) depicts the average linear speed of riblets over time. Consistent with the initial length, the average speed of all riblets at each time step does not exhibit any clear temporal trend. The time‐series representation of the average speed here facilitates the identification and removal of outliers. Moreover, inspection of the series reveals that the instantaneous average speed remains in close agreement with the distribution of values shown in Figure~\ref{fig:hist} (e).

Figure \ref{fig:ts} (c) presents the absolute values of average acceleration of riblets as a function of time. This acceleration is obtained from second-order polynomial fits to each riblet's evolution profile. At each time step, the average acceleration of all riblets is calculated and plotted, with the results indicating no systematic temporal variation.

Finally, figure \ref{fig:ts} (d) illustrates the total number of riblets detected at each time step. This count represents the number of riblets visible in each frame of the observation sequence. The first non-zero count occurs at 12{:}41{:}42~UTC, which we adopt (to within one frame of the CRISP cadence) as the onset time of riblet formation for cross-instrument comparison. Although the event count shows distinct dips at certain times, these variations are attributed to frames with poor seeing conditions rather than any underlying physical process and hence, no definitive physical interpretation can be ascribed to these fluctuations.

\begin{figure}
  \centering
  \includegraphics[width=0.5\textwidth]{ 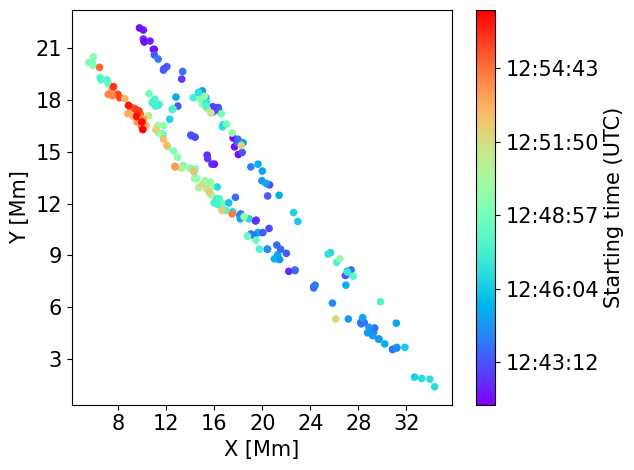}
  \caption{Temporal evolution of the spatial distribution of riblets }
  \label{fig:3d_2dplot}
\end{figure}

We also investigate the temporal evolution of riblet spatial distribution during flare progression. As shown in Figure~\ref{fig:3d_2dplot}, riblets are broadly distributed along the ribbon throughout the observation. The time–color encoding in Figure~\ref{fig:3d_2dplot} also traces the ribbon’s build-up: along the outer ribbon strand, riblets appear first near the bottom right of the field and then progress toward the top left, whereas along the inner strand the prevailing trend is the opposite (top left to bottom right), aside from a few outliers. This strand-wise progression, captured by the systematic color changes, reflects the sweeping motion of the ribbon fronts and complements the overall lack of strong spatial bias. While this suggests a coherent spatio-temporal ordering along each strand, the inference is drawn from a single flare. To determine whether this behavior is a general feature, similar analyses across a larger sample of flares are required.

\subsection{UV and EUV Spectral properties of riblets} \label{aia_lc}

To further elucidate the flare’s temporal evolution, we analyzed AIA lightcurves for the flare.  Lightcurves were extracted from a field of view encompassing both the flare ribbons and associated riblet activity. The intensity values for each wavelength were spatially averaged and the time series was smoothed by fitting a polynomial to reduce noise, and normalized between 0 and 1 for direct comparison between different AIA passbands. For each smoothed lightcurve, two key temporal markers were identified,
Start of Rise: The time the lightcurve enters its sustained rising phase and
Maximum Peak: The time when the normalized lightcurve reaches its highest value.

\begin{figure}
    \centering
    \includegraphics[width=0.5\textwidth]{ 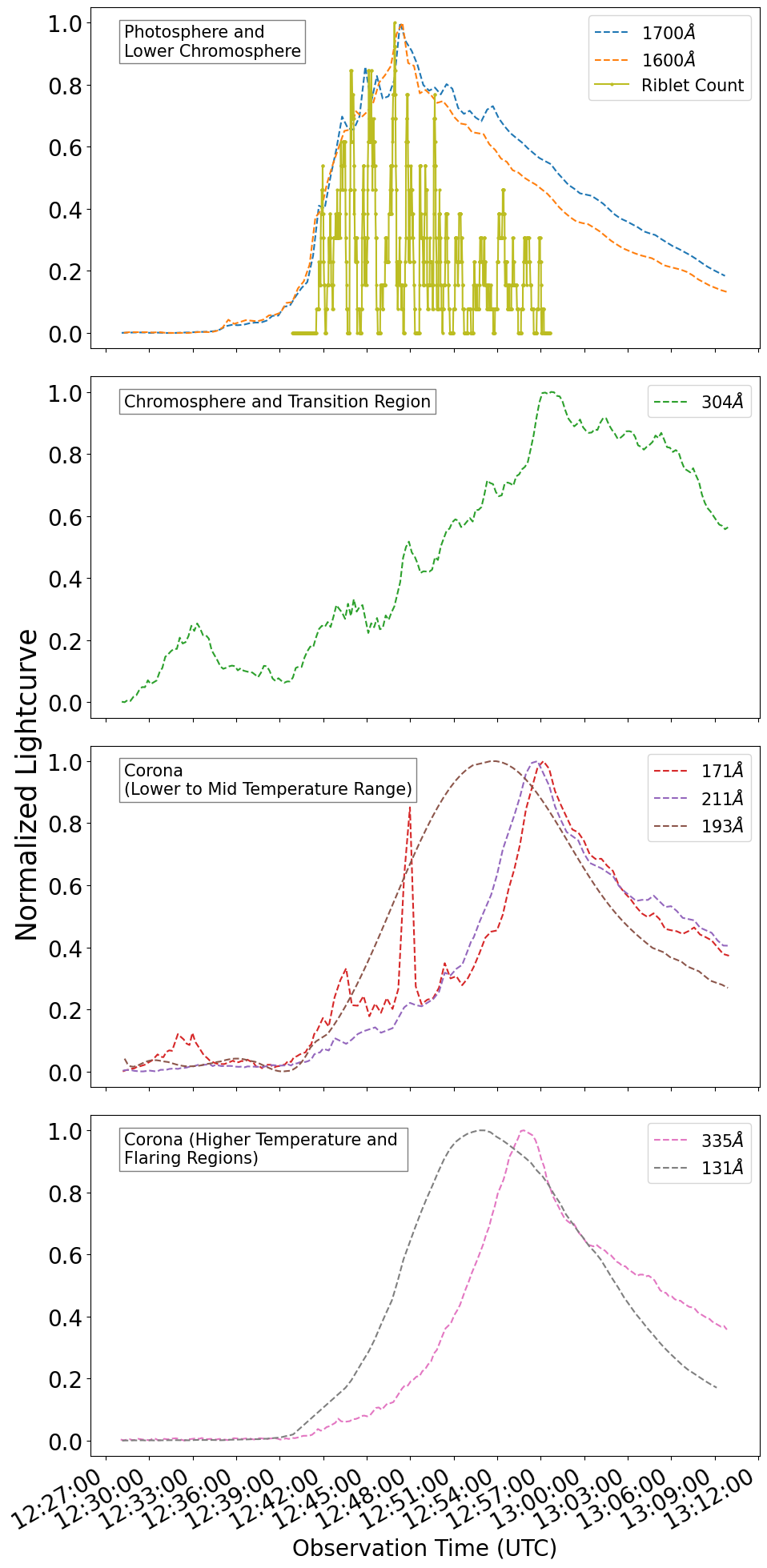}
    \caption{Normalised AIA lightcurves grouped by spectral line formation with respect to solar atmospheric layers and event count of riblets.}
    \label{fig:aia_grouped}
\end{figure}

The timings for these markers are listed in Table \ref{tab:combined_sorted}. Figure \ref{fig:aia_grouped} presents the same lightcurves grouped by the solar atmospheric layers they sample. The overlaid riblet event count reveals that the peak number of riblets is near coincident (within $\lesssim$1 min) with the 1600 \AA{} and 1700 \AA{} maxima, while the coronal channels peak after the riblet count but before the 304 \AA{} channel peak.

Physically, this sequence supports the standard flare model. The initial rise in the 1600~\AA{} and 1700~\AA{} channels corresponds to the onset of magnetic reconnection in the corona, which impulsively accelerates electrons downward along newly reconnected field lines. These electrons deposit their energy predominantly in the chromosphere, producing prompt UV continuum and line emission that manifests as flare ribbon brightenings in the 1600~\AA{} and 1700~\AA{} passbands.
Riblets, small scale, transient brightenings along the ribbon front, emerge concurrently with this UV peak.Their event rate reaches a maximum during the impulsive UV ribbon phase and close to the HXR/GOES-derivative peak times, suggesting that riblet dynamics trace fine structuring of energy deposition at flare footpoints.
This is followed by coronal heating and evaporation (131~\AA{}, 193~\AA{}, 171~\AA{}, 335~\AA{}, 211~\AA{}), as continued energy release drives chromospheric evaporation, filling newly reconnected loops with hot plasma. This plasma emits in the high temperature AIA channels, with peak intensities observed sequentially in cooler coronal lines as the loops radiatively cool.
The transition region response is also evident in 304~\AA{}. While the 304~\AA{} lightcurve begins rising relatively early, it reaches the latest observed maximum, consistent with continued transition region emission as heated plasma evolves and cools.

Thus, the observed riblets occupy the critical temporal window immediately following the first UV ribbon brightenings and just before the coronal emission peaks.  They provide a high resolution diagnostic of nonthermal energy deposition and magnetic footpoint fragmentation during the impulsive phase of flares. Part of the increased emission of transition region lines can be due to transient ionization \citep{doyle_diagnostic_2012}. These lines can be enhanced in the first fraction of a second with the peak in the line contribution function occurring initially at a higher electron temperature than in ionization equilibrium. The rise time and enhancement factor depends mostly on the electron density. The fractional increase in the emissivity due to transient ionization can reach a factor of $\sim$ 2–4.

\begin{table*}
\centering
\begin{tabular}{|l|l|l|l|}
\hline
\multicolumn{2}{|c|}{Sorted by Rising Phase} & \multicolumn{2}{c|}{Sorted by Maximum Peak} \\ \hline
Instrument / Channel & Rising Phase & Instrument / Channel & Maximum Peak \\ \hline
AIA 1700 \AA{} & 12:40:30 & RHESSI 50--100 keV & 12:43:00 \\ \hline
AIA 1600 \AA{} & 12:40:40 & RHESSI 25--50 keV & 12:43:12 \\ \hline
CRISP H$\alpha$ & 12:41:53 & CRISP H$\alpha$ & 12:45:40 \\ \hline
Riblet Count & 12:42:15 & Riblet Count & 12:46:55 \\ \hline
GOES 0.5--4 \AA{} Derivative & 12:42:37 & GOES 0.5--4 \AA{} Derivative & 12:47:04 \\ \hline
RHESSI 25--50 keV & 12:42:44 & GOES 1--8 \AA{} Derivative & 12:47:18 \\ \hline
RHESSI 50--100 keV & 12:42:44 & AIA 1700 \AA{} & 12:47:18 \\ \hline
AIA 171 \AA{} & 12:42:47 & FERMI (25--50 keV) & 12:47:20 \\ \hline
GOES 1--8 \AA{} Derivative & 12:42:50 & FERMI (50--100 keV) & 12:47:23 \\ \hline
FERMI (50--100 keV) & 12:43:01 & AIA 1600 \AA{} & 12:47:28 \\ \hline
AIA 193 \AA{} & 12:44:18 & GOES 0.5--4 \AA{} & 12:50:26 \\ \hline
AIA 304 \AA{} & 12:44:19 & GOES 1--8 \AA{} & 12:52:23 \\ \hline
AIA 131 \AA{} & 12:44:44 & AIA 131 \AA{} & 12:52:44 \\ \hline
FERMI (25--50 keV) & 12:44:48 & AIA 193 \AA{} & 12:53:54 \\ \hline
GOES 0.5--4 \AA{} & 12:47:10 & AIA 335 \AA{} & 12:55:50 \\ \hline
GOES 1--8 \AA{} & 12:48:44 & AIA 211 \AA{} & 12:56:47 \\ \hline
AIA 335 \AA{} & 12:48:26 & AIA 171 \AA{} & 12:57:11 \\ \hline
AIA 211 \AA{} & 12:48:47 & AIA 304 \AA{} & 12:57:43 \\ \hline
\end{tabular}
\caption{Timing of key features observed in CRISP H$\alpha$, riblet count, FERMI, RHESSI, GOES, GOES derivatives, and AIA light curves. Left side sorted by rising phase, right side sorted by maximum peak. All times in UTC.}
\label{tab:combined_sorted}
\end{table*}

To provide additional context for the temporal evolution of the  CRISP H$\alpha$ lightcurve, we incorporated lightcurves from Fermi, RHESSI, and GOES during the observed flare period (Figure~\ref{fig:images}c). These were analyzed to identify the onset of the rising phase and the peak time, following the same approach as for the AIA lightcurves. The corresponding timings are listed in Table~\ref{tab:combined_sorted}.

The earliest rising trends were observed in the AIA 1700~\AA{} and 1600~\AA{} channels, followed by increases in the CRISP H$\alpha$ intensity and the riblet event count ($\sim$1--2 min after the initial UV rise). The HXR channels (RHESSI 25--50 and 50--100~keV, followed by Fermi) enter their sustained rising phase shortly thereafter, while the latest increases occur in the GOES SXR channels.

In terms of peak timing, RHESSI 50--100~keV reached its maximum first, though its coverage was cut short due to the satellite entering eclipse (Earth occultation). The CRISP Halpha lightcurve peaked before the riblet event count, and remained elevated for an extended duration. Fermi and the GOES derivatives peaked nearly simultaneously, followed by the GOES 0.5--4~\AA{} and then the 1--8~\AA{} SXR channels.
The close temporal alignment between the riblet event peak and the HXR/GOES-derivative peaks suggests that riblet brightness is driven by impulsive non-thermal energy deposition rather than gradual thermal conduction.

The temporal correspondence between the GOES SXR derivative and RHESSI 50--100~keV confirms a standard Neupert effect pattern, with nonthermal electron precipitation driving both diagnostics. The HXR rise and peak times in RHESSI/Fermi occur during the same impulsive interval as the enhanced CRISP H$\alpha$ emission and riblet activity, together tracing rapid chromospheric heating and fine-scale ribbon structuring. The delayed GOES SXR maxima reflect gradual plasma filling and cooling in reconnected coronal loops during the flare decay. The extended plateau in CRISP intensity and sustained riblet count indicate prolonged energy release and ribbon evolution in the lower atmosphere.

\subsection{X-ray Spectral properties of riblets}\label{sec:xray}

\begin{figure*}[t]
  \centering

  \begin{minipage}[t]{0.59\textwidth}
    \vspace{0pt}
    \centering
    \includegraphics[width=\textwidth]{ 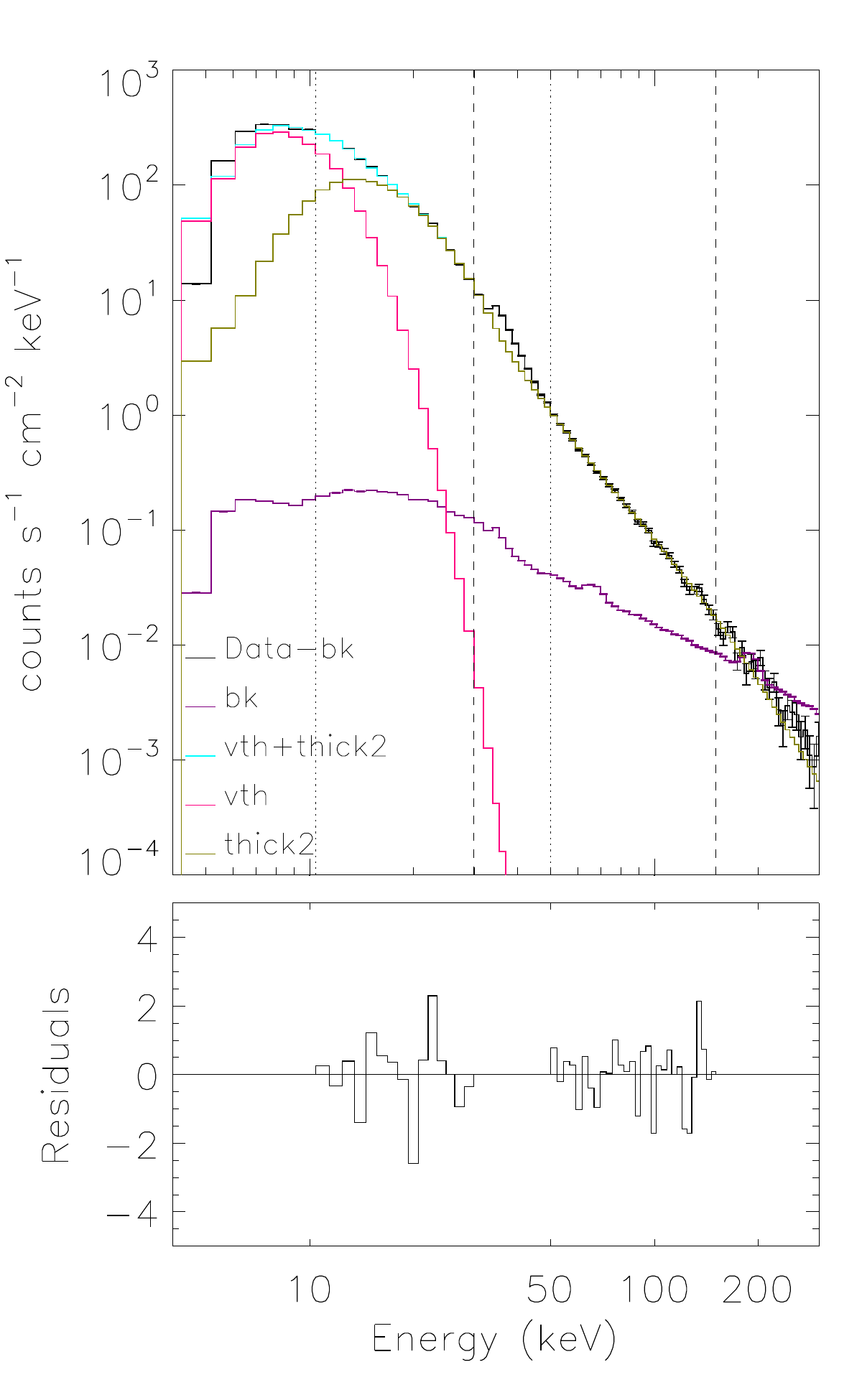}
  \end{minipage}
  \hfill
  \begin{minipage}[t]{0.39\textwidth}
    \vspace{0pt}
    \centering
    \includegraphics[width=\textwidth]{ 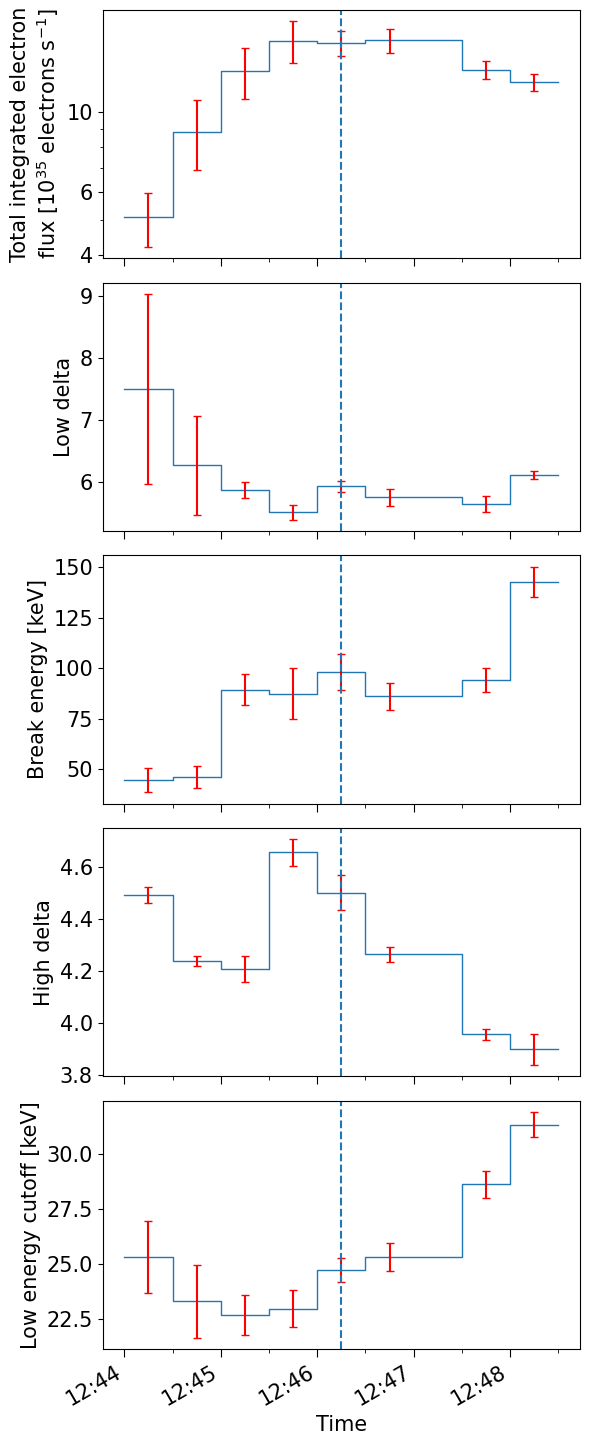}
  \end{minipage}

  \caption{Fermi/GBM spectral analysis near the riblet peak.
  Top left panel: Count spectrum with best-fit \texttt{vth+thick2} model for
  12{:}46{:}00.355–12{:}46{:}30.052~UT. Vertical dashed lines mark the fitted cutoff/break energies.
  Bottom left panel: Fit residuals for the same time interval.
Right panel: Time evolution of selected nonthermal fit parameters from successive 30s spectral fits with uncertainties plotted at the bin midpoints (red) and connected as a step function across each integration bin. The blue vertical dashed line marks the midpoint of the interval used for the spectral fit and for subsequent calculations.  }
  \label{fig:gbm_fit_resid}
\end{figure*}

We model the Fermi GBM spectra with an isothermal component plus a collisional thick–target power law (\texttt{vth+thick2}). The example fit for 12{:}46{:}00.355–12{:}46{:}30.052~UT is shown in Fig.~\ref{fig:gbm_fit_resid}(top left panel). The residuals in Fig.~\ref{fig:gbm_fit_resid} (bottom left panel) are approximately symmetric, and in a non-systematic distribution about zero across the fitted range, indicating that the model provides an adequate description of the data near the time of interest.
Using consistent OSPEX settings over the interval 12{:}39–12{:}58~UT, we extract the total integrated electron flux $\dot N_e$ (electrons~s$^{-1}$), the low-energy spectral index $\delta_{\rm low}$, and the low-energy cutoff $E_{\rm c}$ shown in Fig.~\ref{fig:gbm_fit_resid} (right panel).

Our OSPEX nonthermal parameters map directly to a standard flare electron-beam injection parameterization used in radiative-hydrodynamic modeling.  We use the total integrated electron rate above $E_{\rm c}$, $\dot N_{e}$ (electrons s$^{-1}$), the power-law spectral index $\delta$, and the low-energy cutoff $E_{\rm c}$ to define a single-power-law injection
\[
F(E)\ \propto\ E^{-\delta}\qquad (E\ge E_{\rm c}).
\]
The mean electron energy $\langle E \rangle$ is defined as
\begin{equation}
    \langle E \rangle =
    \frac{\displaystyle \int_{E_{\rm c}}^{\infty} E \, F(E)\,{\rm d}E}
         {\displaystyle \int_{E_{\rm c}}^{\infty} F(E)\,{\rm d}E}.
\end{equation}
For $\delta>2$ this evaluates to
\[
\langle E\rangle=\frac{\delta-1}{\delta-2}\,E_{\rm c},
\]
so the instantaneous power carried by the nonthermal electrons is
$P=\dot N_{e}\,\langle E\rangle$.

The fitted thick-target component is parameterized with a break energy $E_b$ and indices $\delta_{\rm low}$ and $\delta_{\rm high}$ (Fig.~\ref{fig:gbm_fit_resid}, right panel). During the riblet-peak interval, $E_{\rm c}\ll E_b$, so for estimating $\langle E\rangle$, $P$, and $\mathcal{F}_{\rm beam}$ we approximate the injected spectrum above $E_{\rm c}$ as a single power law with index $\delta\equiv\delta_{\rm low}$.

 The corresponding electron–beam energy flux is
\[
\mathcal{F}_{\rm beam}=\frac{P}{A_{\rm fp}},
\]
where $A_{\rm fp}$ is a characteristic flare footpoint area.

We estimate $A_{\rm fp}$ from RHESSI CLEAN images in the 25--50~keV band. Using the 50~\% of maximum intensity contour of the brightest compact kernel as the characteristic precipitation area, we obtain $A_{\rm fp}\approx 5.19\times 10^{18}$~cm$^{2}$. At the local riblet peak (12{:}46{:}15~UT bin midpoint), the OSPEX parameters are $\dot N_{e}(>E_{\rm c})=(15.62\pm 1.27)\times 10^{35}$ electrons~s$^{-1}$, $\delta_{\rm low}=5.93\pm 0.09$, $E_{\rm c}=24.72\pm 0.54$~keV, and $E_b=98.2\pm 8.9$~keV. Using $\delta\equiv\delta_{\rm low}$, we find $\langle E\rangle = (31.0\pm 0.7)$~keV and hence $P=(7.76\pm 0.65)\times 10^{28}$~erg~s$^{-1}$. This gives a characteristic beam energy flux
\[
\mathcal{F}_{\rm beam} = \frac{P}{A_{\rm fp}} \approx (1.50\pm 0.13)\times 10^{10}\ {\rm erg\ cm^{-2}\ s^{-1}},
\]
consistent with an F10-class electron beam in the sense of commonly used RADYN flare runs.
Having established representative nonthermal electron-beam parameters from the Fermi/GBM spectroscopy, we next test whether riblet properties exhibit spatial organization along the ribbon. In particular, we examine whether linear and non-linear riblets preferentially occur in distinct locations, or cluster in regions suggestive of locally enhanced energy deposition.

\subsection{Spatial properties of riblets}

\begin{figure*}
  \centering
  \subfloat{
    \includegraphics[width=0.75\linewidth]{ 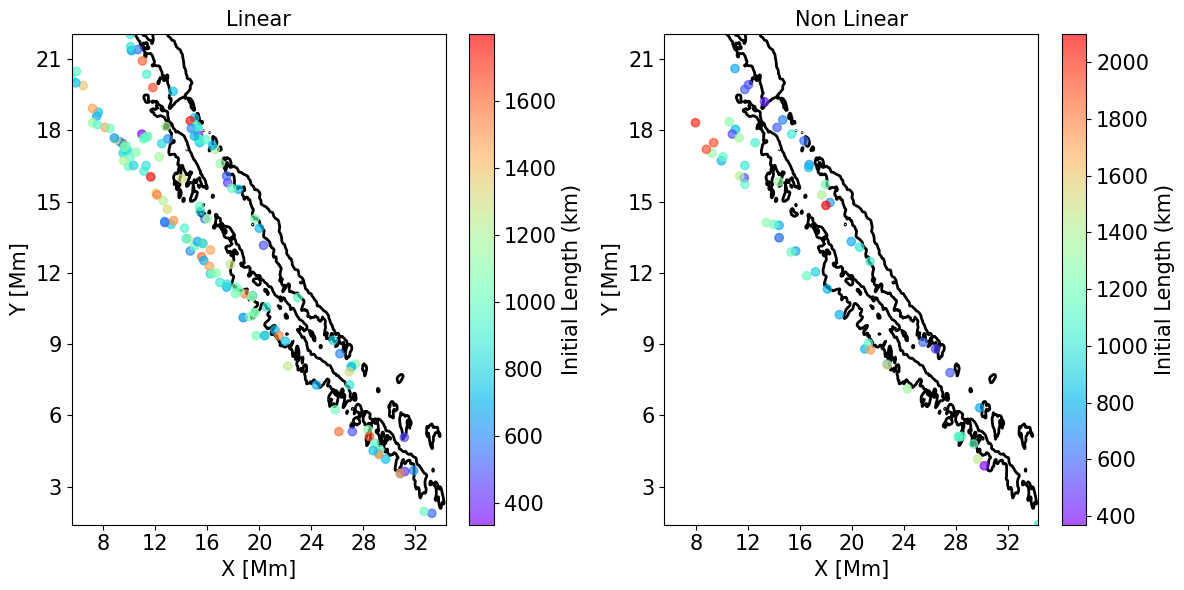}

  }\\[\baselineskip]
  \subfloat{
    \includegraphics[width=0.75\linewidth]{ 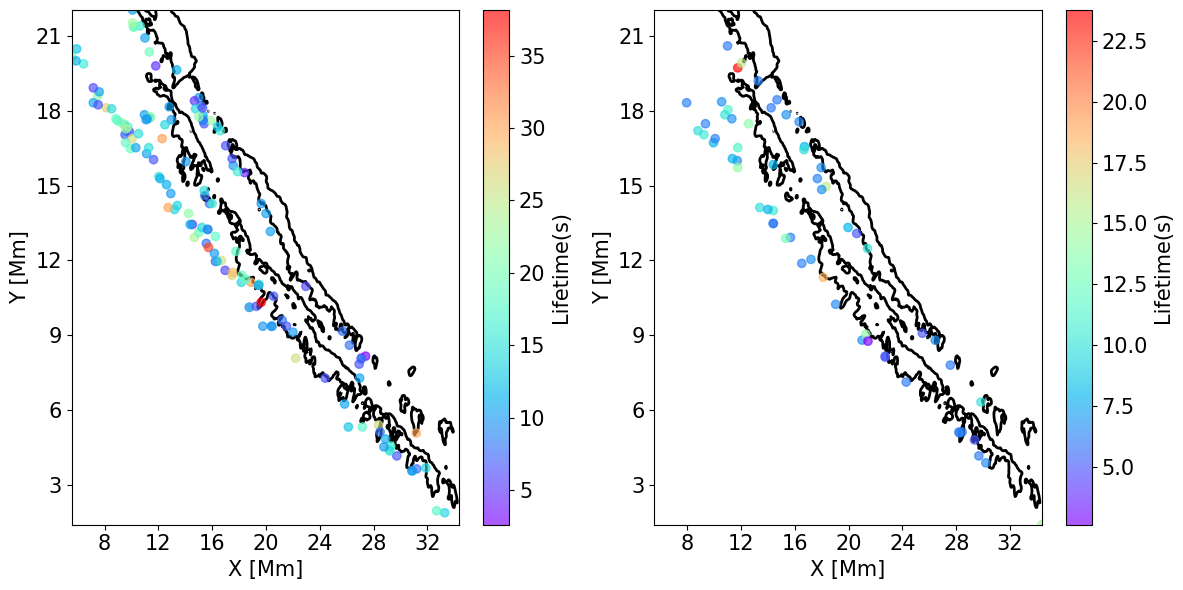}

  }\\[\baselineskip]
  \subfloat{
    \includegraphics[width=0.75\linewidth]{ 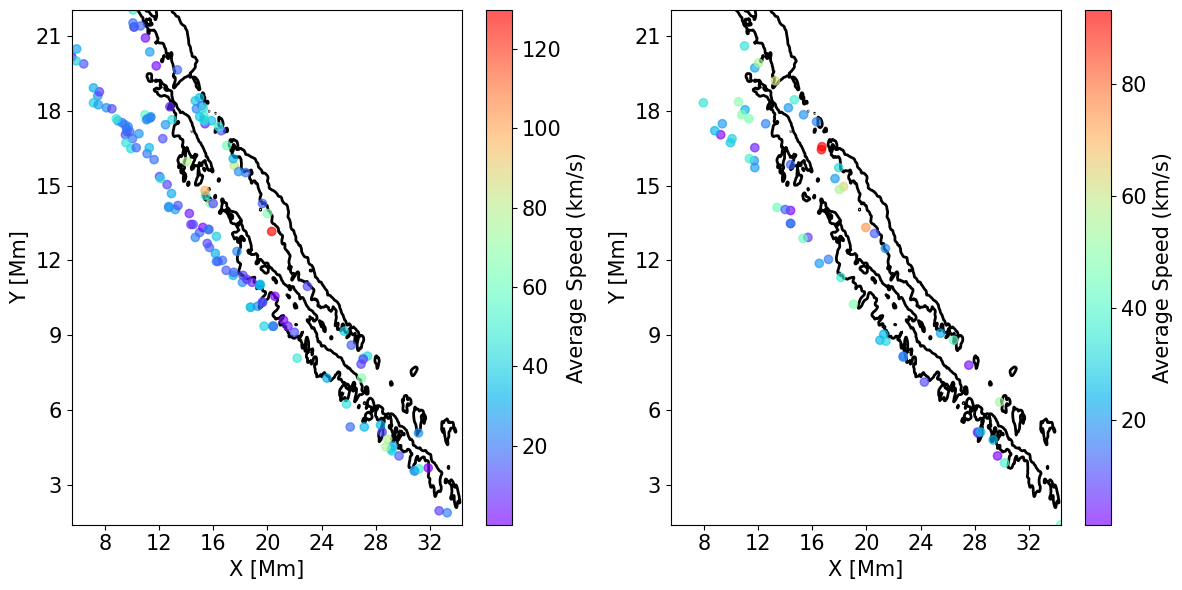}
  }
  \caption{Spatial distribution of initial length, lifetime, and average speed of riblets over the entire duration of the observation. The left column corresponds to linear cases and the right column to non-linear cases. The black contour represents the position of the ribbon in the field of view at time 12:51:53 UTC.}
  \label{fig:sp}
\end{figure*}

Figure~\ref{fig:sp} shows the spatial distribution of key riblet parameters across the solar surface. The left and right columns correspond to linear and non-linear riblets, respectively, and illustrate their initial length, lifetime, and average speed.

Examination of the spatial distributions reveals that all three parameters are uniformly spread across the entire ribbon for both linear and non-linear cases. One notable observation is that riblets with longer lifetimes and larger initial lengths are predominantly located on the portion of the ribbon closer to the limb. This spatial bias can be attributed to the fact that riblets emerging from regions further from the limb, if sufficiently long, tend to merge into the ribbon segment nearer to the limb from our viewpoint, thereby eluding tracking. This can be seen in Figure \ref{fig:images} (b) which shows all the riblet tracks. Additionally, the distribution of average speed shows no significant clustering of higher or lower speeds in either case. Consequently, this analysis does not provide a means to meaningfully differentiate between linear and non-linear riblet cases based on their spatial distribution.

We next examined the riblet spatial distribution as a function of local energy release by overlaying  CRISP H$\alpha$ observations with contours from both AIA and RHESSI.

\begin{figure*}
  \centering
  \subfloat[]{%
    \includegraphics[width=0.49\linewidth]{ 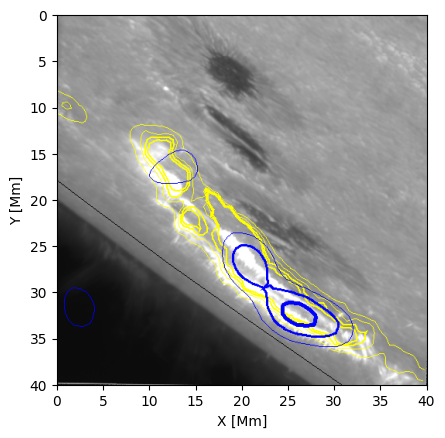}%
    \label{fig:1700_rhessi}%
  }\quad
  \subfloat[]{%
    \includegraphics[width=0.49\linewidth]{ 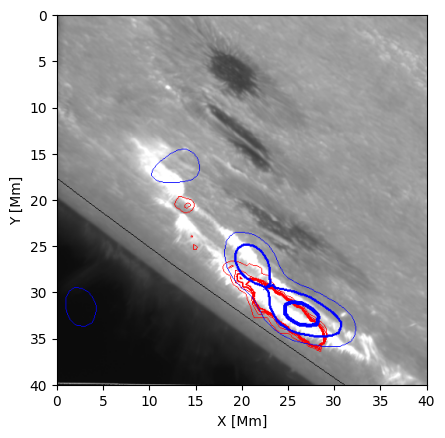}%
    \label{fig:304_rhessi}%
  }

  \caption{CRISP image (Time- 12:51:53 UTC) overlaid with (a) AIA 1700~\AA{} (yellow) intensity contours and (b) AIA 304~\AA{} (red) and RHESSI 25–50 keV X-ray contours (blue in both (a) and (b) ). The RHESSI loop-top source is evident at approximately $(2.5,31.5)$ Mm. }
  \label{fig:aia_rhessi_contours}
\end{figure*}

Figure~\ref{fig:aia_rhessi_contours} (b) shows the AIA 304 \AA{}, which samples plasma at transition region temperatures (around 50,000K), with RHESSI 25–50keV contours indicating sites where nonthermal electrons deposit energy via thick-target bremsstrahlung.  The RHESSI contour positioned at $(2.5,31.5)$ Mm, corresponds to the flare loop-top region. RHESSI images in the 25--50~keV range were integrated over the interval 12:38:00--12:43:30~UTC to generate a single contour map. This approach was necessary because RHESSI coverage spans only a short period before the spacecraft entered eclipse, and shorter integration windows produced low count rates and large uncertainties. As a result, the RHESSI contours remain fixed throughout our analysis. 

In contrast, AIA provides a time sequence of images, with intensities evolving significantly from frame to frame, albeit at a lower cadence as compared with CRISP. Consequently, the location of a given contour level (e.g., 40\%) in one snapshot may not correspond to the same relative brightness in another. To establish a consistent reference, we therefore analyzed the intensity distribution across the entire field of view and time span of the observation for both the AIA 304~\AA{} and 1700~\AA{} channels. Contour levels were then defined as fixed percentages (e.g., 10\%, 20\%, etc.) of this global intensity distribution, ensuring that riblet positions could be compared against a stable set of reference levels despite the evolving AIA emission.

Figure~\ref{fig:aia_rhessi_contours} (a) shows the AIA 1700~\AA{} channel, dominated during flares by chromospheric emission, with RHESSI 25--50\,keV contours overlaid. The spatial coincidence of RHESSI footpoints with enhancements in both AIA 304~\AA{} and 1700~\AA{} implies a common driver precipitating nonthermal electrons depositing energy across multiple atmospheric layers.

\begin{figure}
  \centering
  \subfloat[]{
    \includegraphics[width=0.45\textwidth]{ 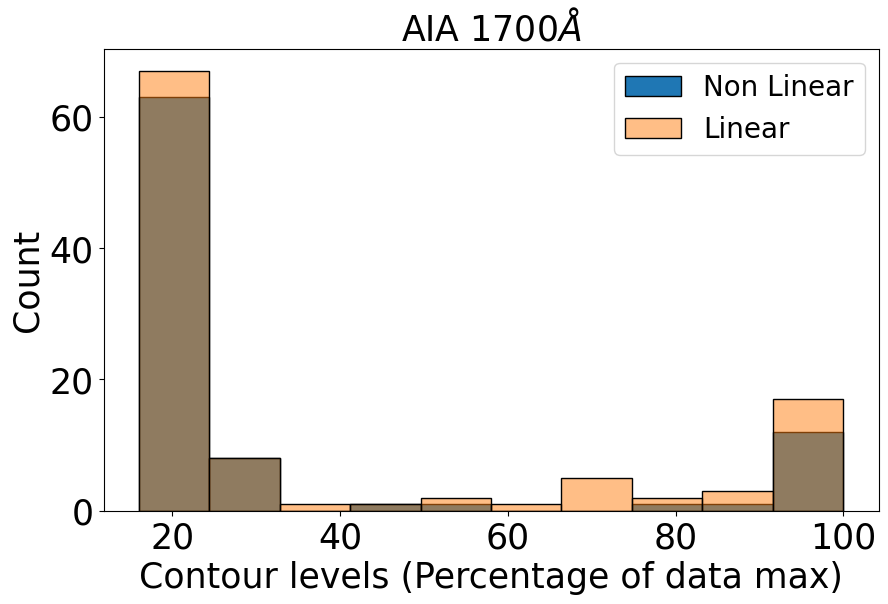}
  }\quad
  \subfloat[]{
    \includegraphics[width=0.45\textwidth]{ 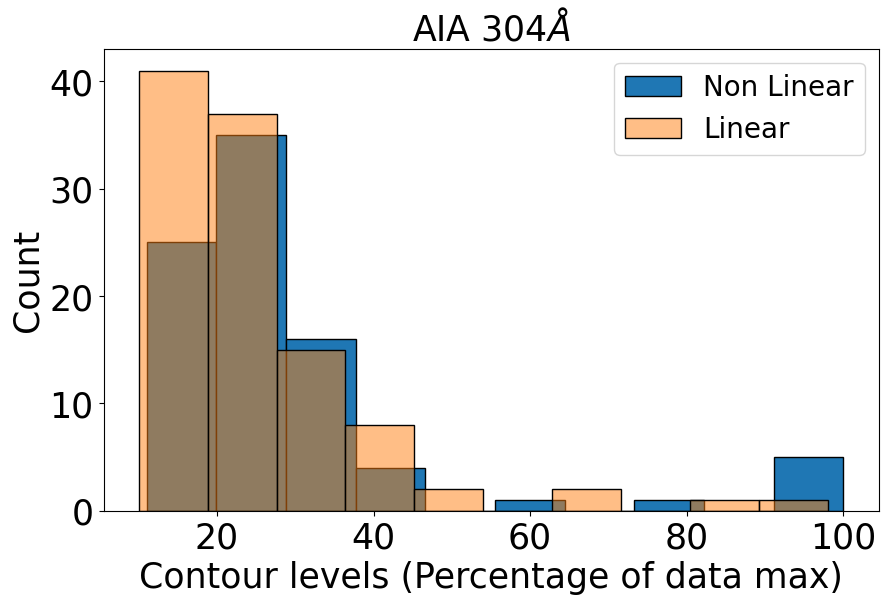}
  }

  \vskip\baselineskip

  \subfloat[]{
    \includegraphics[width=0.45\textwidth]{ 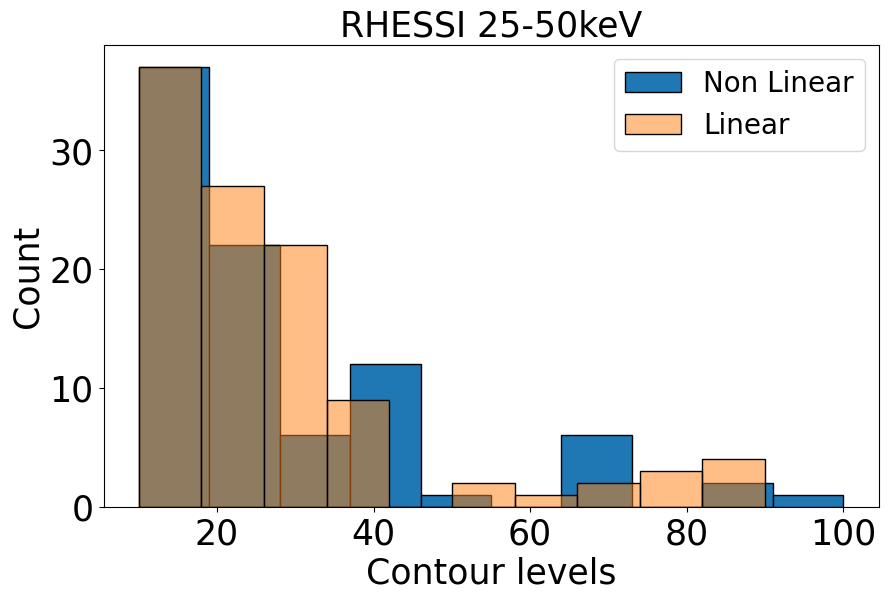}
  }

  \caption{Histograms of riblet occurrence at different contour levels in 
   \textit{(a)} AIA 1700\AA{}, 
   \textit{(b)} AIA 304\AA{}, and 
   \textit{(c)} RHESSI 25–50 keV. In (a) and (b) the x-axis shows contour levels expressed as a percentage of each frame’s intensity maximum (levels recomputed for every AIA image), whereas in (c) the x-axis shows absolute contour levels from a single RHESSI image integrated over the full observing interval.
  }
  \label{fig:contour_hist}
\end{figure}

Figure~\ref{fig:contour_hist} quantifies riblet occurrence within successive percentile levels of the AIA 1700~\AA{}, AIA 304~\AA{}, and RHESSI 25--50~keV intensity distributions. To test the hypothesis that stronger energy deposition favors non-linear evolution, this plot was designed to examine whether non-linear riblets preferentially co-locate with higher intensity regions and to quantify their correspondence with the brightest emission across independent diagnostics.
The overlapping histograms for linear and non-linear riblets show no systematic differences between the two classes. Moreover, riblets occur across the full range of AIA and RHESSI intensity levels, indicating no clear preference for regions of enhanced emission. The fact that most riblets are associated with $<40\% $ contour levels implies that there is a tentative correspondence between the brightest emissions in the respective wavelength passbands. But not a strong coincidence. This suggests that riblet dynamics are not strongly governed by the local energy deposition traced by these diagnostics. However, because RHESSI coverage was limited and only a single integrated contour could be used, a time-resolved comparison between riblet evolution and HXR footpoints remains for future investigation.

\section{Towards understanding the linear and non-linear difference} \label{geometry}

At the conclusion of the analysis in Section~\ref{sec:stats}, we find no observationally robust basis for distinguishing linear from non-linear riblets. The spatio-temporal distributions of length, lifetime, and speed for the two classes substantially overlap. Likewise, occurrence as a function of AIA~1700~\AA{}, AIA~304~\AA{}, and RHESSI~25--50~keV intensity percentiles shows no enhancement of one class over the other. Thus, using these observables, we do not obtain a statistically reliable discriminator between linear and non-linear riblets. In order to shed more light on the reason for the classifications we further investigate the riblets' morphology, particularly the potential effects on the kinematics due to the geometrical shapes and curvature of the riblet projections relative to the line of sight.

\begin{figure*}
  \centering
  \subfloat[]{
    \includegraphics[width=0.45\textwidth]{ 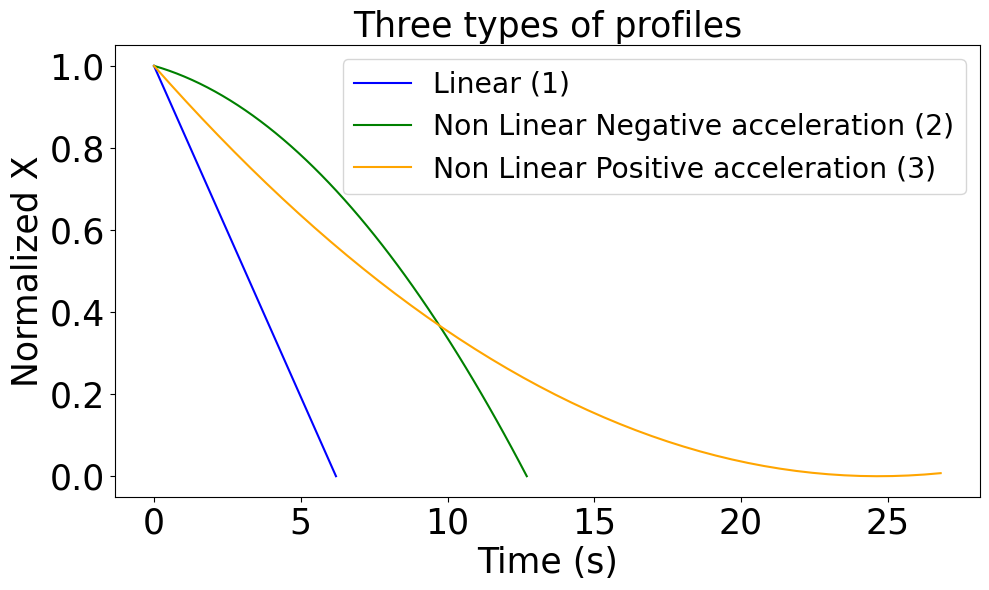}
    \label{fig:profiles}
  }\quad
  \subfloat[]{
    \includegraphics[width=0.45\textwidth]{ 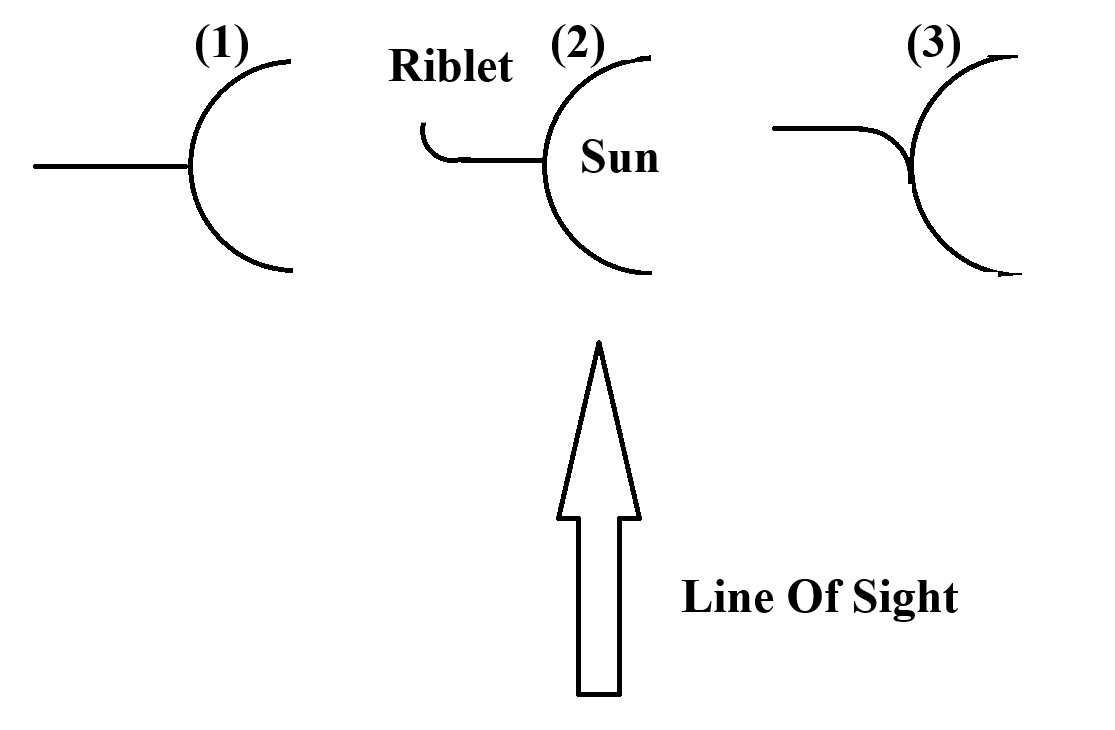}
  }

  \caption{
   \textit{(a)} Three different types of observed profiles: linear, non-linear with negative acceleration, and non-linear with positive acceleration. 
   \textit{(b)} Three different curvatures corresponding to the profiles: (1) linear (constant speed), (2) non-linear with negative acceleration, (3) non-linear with positive acceleration.
  }
  \label{fig:curvatures}
\end{figure*}

As illustrated in Figure~\ref{fig:curvatures} (a), riblets exhibit three distinct types of apparent speed evolution: (1) constant speed (linear), (2) deceleration (non-linear with positive acceleration), and (3) acceleration (non-linear with negative acceleration). These profiles summarize the range of kinematic behaviours identified in our sample before considering projection effects. Because riblet length decreases with time ($\Delta x<0$), the sign of the fitted acceleration refers to changes in the magnitude of the shortening speed: $a>0$ means speed decreases (profile flattens in $X$–$T$ plots), whereas $a<0$ means speed increases (profile steepens).

We propose three different structural interpretations for these profiles to account for positive and negative acceleration interpretations as derived from X-T plots.
As shown in Figure \ref{fig:curvatures} (b), the first profile represents a riblet perfectly perpendicular to the line of sight which could be envisaged as a perfectly radial projection of a riblet at the limb of the Sun, which evolves linearly in our observations. The second curvature corresponds to the hypothetical scenario for the geometry of a non-linear profile with negative acceleration. In this case, as the riblet decreases in length and becomes more aligned with the line of sight, we initially observe minimal motion in the plane of sky, followed by an increase in apparent motion as the riblet's length shortens. The third curvature represents the hypothetical scenario of a non-linear profile with positive acceleration. The riblet starts perpendicular to the line of sight and, as it decreases in length, becomes more parallel, leading to a reduced apparent speed in the plane of sky.

These proposed projected profiles imply logically that riblets are not confined to a single, uniform orientation but instead trace paths with varying curvature and inclination. If riblets indeed follow magnetic field lines, the diversity of apparent geometries indicates that the local magnetic field at the ribbon front is structured and non-uniform, with different strands reconnecting and retracting along slightly different trajectories. Such variation is consistent with fine scale magnetic complexity at flare footpoints, as inferred from earlier observations of sub-arcsecond kernels and current mapping into ribbon fine structures \citep[e.g.,][]{Janvier2016,Brannon2015,2024PietrowRibbon}.

\begin{figure}
  \centering
  \subfloat[]{
    \includegraphics[width=0.49\textwidth]{ 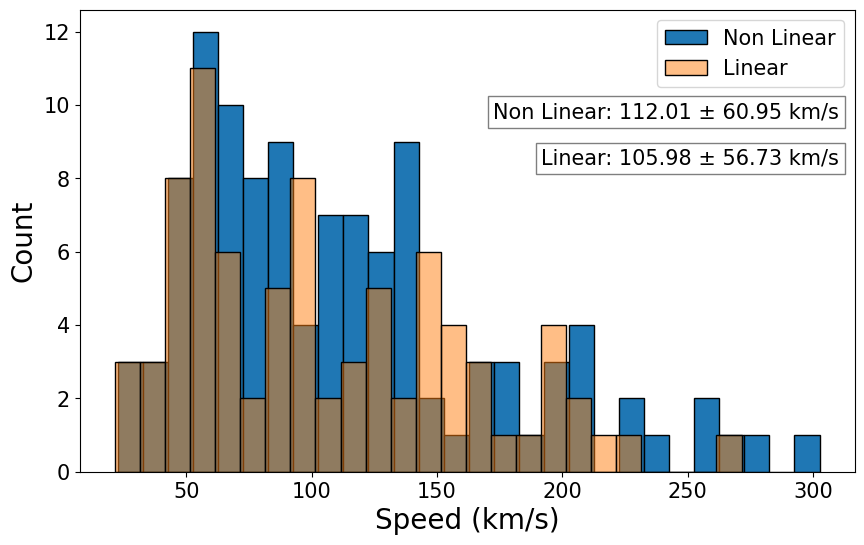}
    \label{fig:speed_curv_hist}
  }\quad
  \subfloat[]{
    \includegraphics[width=0.49\textwidth]{ 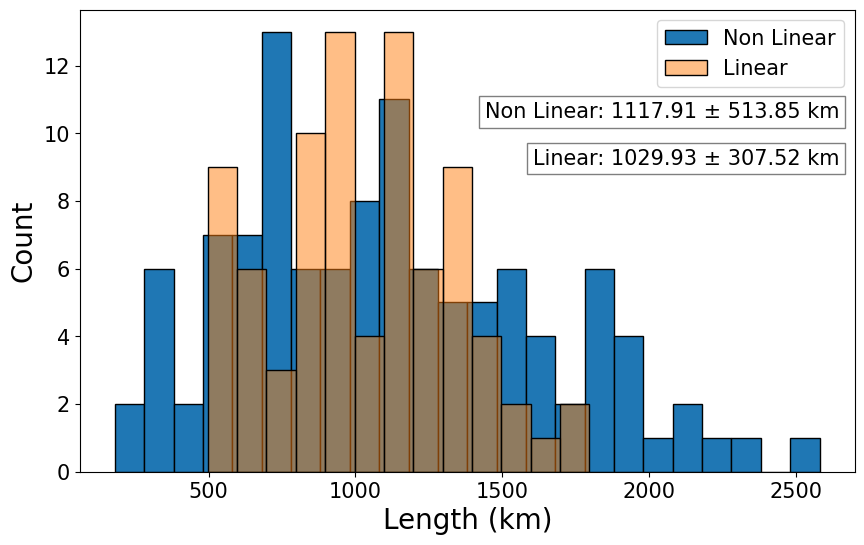}
    \label{fig:len_curv_hist}
  }
  \caption{Histograms of \textit{(a)} updated speed and \textit{(b)} initial length of the riblets, accounting for curvature.}
  \label{fig:curv_hist}
\end{figure}

Assuming that all riblets evolve linearly, including the observed non-linear cases, we can consider the initial speed for the profile with positive acceleration and the final speed for the profile with negative acceleration as the true speeds of the riblets and since in this assumption the speed does not change during the lifetime of the riblets, we used these speeds and the observed time, to derive updated lengths for the riblets. Figure \ref{fig:curv_hist} presents histograms of the linear and ``non linear" cases with the updated lengths and speeds. This analysis indicates that riblets  with curvature are generally longer and faster (average length of 1117.91 km and average speed of 112.01 km/s), which is consistent with projection geometry: length of curved riblets corrected for projection effects are longer while their measured lifetimes are unchanged, thereby yielding higher average velocities. Despite this analysis, we still lack a clear physical explanation for the observed variations in riblet classifications. The next step should be to utilize numerical modeling of electron beam injections into different chromospheric atmospheres, in order to attempt to recreate these riblet evolutions and better understand the reasons behind these different properties for these classes. In particular, the beam parameters derived from the Fermi/GBM analysis (Sec.~\ref{sec:xray}) $\dot N_{e}$, $\delta$, $E_{\rm c}$, and $\mathcal{F}_{\rm beam}$ provide a compact set of  inputs for RADYN-type simulations aimed at reproducing riblet-like substructures.

\section{Discussion}\label{sec:disc}

The presence of riblets with both linear and non-linear evolution patterns highlights the complexity of flare‐ribbon dynamics. The spatial and temporal resolution of our observations, including continuous coverage through the rise phase of an X1.5 limb flare, enabled frame-by-frame tracking of large riblet samples and a statistical comparison of their properties.

Riblets appear intermittently during the impulsive phase, both spatially and temporally. In Section~\ref{sec:temp_dist} we showed that riblets are distributed along the ribbon (Fig.~\ref{fig:3d_2dplot}) without an obvious concentration at particular footpoints within our field of view. We identified two classes of apparent evolution, constant speed (``linear'') and varying speed (``non-linear''). Despite the difference in kinematic profiles, their initial speeds, lengths, and lifetimes are not statistically distinguishable (Sec.~\ref{sec:stats}, Fig.~\ref{fig:hist}). As discussed in Section~\ref{geometry}, projection can plausibly contribute to the apparent dichotomy, but we do not attempt a quantitative de-projection here.

The spatial comparison with AIA and RHESSI diagnostics (Fig.~\ref{fig:contour_hist}) shows riblets across the full range of AIA\,1700~\AA{}, AIA\,304~\AA{}, and RHESSI 25–50~keV intensities, with no clear preference for the brightest regions. Overlapping histograms for linear and non-linear riblets indicate that the two classes remain statistically indistinguishable in these diagnostics. A more definitive assessment of occurrence relative to HXR footpoints will require time-resolved imaging spectroscopy to overcome the limitations of static RHESSI contours.

AIA lightcurves (Sec.~\ref{aia_lc}, Fig.~\ref{fig:aia_grouped}) place the riblets within the impulsive phase: the riblet count peaks with AIA\,1600~\AA{} and 1700~\AA{} maxima and the GOES derivative and FERMI/GBM peak, and declines by 12:52~UTC near the GOES soft X-ray maximum. This timing is consistent with a Neupert-type relationship between energy deposition and riblet occurrence. Beyond timing, the Fermi/GBM spectral fits in Sec.~\ref{sec:xray} provide a quantitative constraint on the nonthermal driver. HXR-inferred electron-beam parameters are routinely used as direct inputs to radiative-hydrodynamic flare models. For example, RADYN calculations have been constrained using RHESSI/GBM-like nonthermal properties to model flare kernels and ribbon heating, and to quantify the sensitivity of the atmospheric response to the spectral index and low-energy cutoff \citep{kowalski_flare_model_obs_2017,Kennedy2015,Allred2015,Simoes2015}. In this context, OSPEX analysis of Fermi/GBM spectra, including joint use with RHESSI, provides a practical route to estimating nonthermal electron numbers and energetics \citep[e.g.][]{James2023}.
Around the local riblet peak, our OSPEX fits imply a strong-beam regime. Using $\delta\equiv\delta_{\rm low}$ and the 50\% contour area of the brightest RHESSI 25--50~keV kernel ($A_{\rm fp}\approx 5.19\times10^{18}$~cm$^{2}$; Sec.~\ref{sec:xray}), we obtain a characteristic beam energy flux $\mathcal{F}_{\rm beam}\approx 1.5\times10^{10}$~erg~cm$^{-2}$~s$^{-1}$, consistent with F10-class inputs commonly adopted in RADYN-type flare runs. Our derived energy flux provides a quantitative link between the high-energy driver and the chromospheric response. The observed speeds are broadly comparable to chromospheric condensation velocities reported in previous X-class flare studies, supporting the interpretation of riblets as localized downflows resulting from strong beam heating.

 In this regime beam heating is expected to dominate the chromospheric energy budget during the riblet phase, making riblets a natural target for radiative--hydrodynamic simulations that test how spatially and temporally inhomogeneous electron beams can reproduce the observed ribbon fine structure.

  While riblets share morphological similarities with Type II spicules, several key distinctions suggest a different physical origin. Specifically, the mean lifetime of riblets (~12 s) is significantly shorter than typical spicule durations (10–100 s). Furthermore, the precise temporal and spatial coincidence between riblet emergence and impulsive HXR bursts (see Figure 1(c)) strongly suggests they are transient chromospheric responses to non-thermal electron precipitation rather than ambient mass ejections.

The analysis of signatures associated with chromospheric downflows observed through the H$\alpha$ line has been the subject of previous investigations. For instance, \citet{druett_beam_2017} reported red-shifted emission in a C-class flare, interpreting flare profiles obtained from observations with the SST resulting from a downward flow with velocities of approximately 45–50 km/s. These inward Doppler velocities were also corroborated by consistency between observed profiles and synthesised profiles using the HYDRO2GEN radiative hydrodynamic code, driven by a 3F10 class electron beam in the modeling setup. This is broadly comparable to the beam-flux regime inferred here around the riblet peak (Sec.~\ref{sec:xray}).
 The same model was also employed in \citet{druett_hydro2gen:_2018} to interpret spectroscopic observations from the 1980s \citep{wuelser_high_1989}, attributing the observed profiles to downflows with higher velocities.  \citet{inchimoto_flare_obs_1984} found flare profiles with strongly red-shifted emission peaking up to +3\AA \ from the line center, exhibiting line asymmetries that gradually drifted back towards the central wavelength over approximately 1 minute. \citet{wuelser_high_1989} reported similar profile asymmetries with enhancements peaking up to +2\AA \ from the line center. Such highly Doppler shifted emissions have recently been reinterpreted as potentially originating from coronal rain sources overlapping with flare ribbons in the line of sight \citep{2023XuExtremeRedWing, 2024PietrowRibbon}. The interpretation of these profiles as a coronal source by \citet{2024PietrowRibbon} avoids the need to invoke an exceptionally deep ingress of chromospheric material into the Sun. Additionally, RADYN models have explored this phenomenon, such as the F11 simulation \citep{allred_radiative_2005} producing short-lived downward shocks with velocities on the order of tens of km/s, resulting in asymmetric H$\alpha$ line profiles with enhancements in the near-red wing. An important difference between HYDRO2GEN and RADYN is the construction of the chromospheric model atmosphere through which the beam propagates and consequently, similar downflow responses can be obtained for different nominal beam fluxes, depending on the assumed atmospheric stratification and radiative-transfer treatment. This sensitivity to the background atmosphere provides a plausible route by which local chromospheric conditions could modulate the observed riblet kinematics, potentially contributing to the apparent diversity of profiles (including the linear and non-linear classes).
 Updated models have yielded even higher velocities of short-lived shocks, reaching up to 100km/s \citep{kowalski_radyn_2015}, with H$\alpha$ line profiles peaking around 1\AA{} from the line center in an X1 solar flare \citep{kowalski_flare_model_obs_2017}.

Considering that the focus of our study is the intensity variations, it is plausible for these variations to arise from changes in either density, temperature, or both. Further spatiotemporal analysis will aid in identifying and understanding the physics underlying riblets and the reasons for the existence of these two distinct classes.

It is crucial to emphasize that the intensity variations we observe can be a result of various physical mechanisms, such as plasma motion, transient ionization and heat transfer through conduction or radiation. However, the timescales we are working with can help us investigate and eliminate these possibilities. In the solar chromosphere, plasma can exhibit a broad range of velocities, influenced by factors such as temperature and magnetic fields. It is important to note that the chromospheric plasma is highly dynamic and subject to continuous changes, resulting in significant variations in speeds over time and across different regions of the chromosphere. \cite{kazachenko_database_2017} compiled a comprehensive database of flare ribbon properties, examining correlations between X-ray flux, ribbon properties, and active region characteristics. They found strong correlations between the peak X-ray flux and the flare ribbon reconnection flux, flare ribbon area, and the fraction of active region flux that undergoes reconnection. \cite{simoes_spectral_2019} overcame spectral limitations by analyzing images from SDO/AIA's broadband ultraviolet filters, focusing on morphology and evolution to gain a better understanding of flare ribbons. They found that AIA UV flare excess emission is of chromospheric origin.

\citet{doyle_diagnosing_2013} demonstrated that transient ionization can amplify the Si IV 1394 Å line intensity by a factor of 2–4 within the opening fraction of a second of a flare. This rapid enhancement occurs because the line's contribution function temporarily shifts its peak to a higher electron temperature than what is expected under standard ionization equilibrium conditions. This followed on from earlier work by \citet{doyle_diagnostic_2012} who showed that the O V 1371\AA{} line is enhanced by the same process. The rise time and enhancement factor depend mostly on the electron density. The fractional increase in the O V 1371\AA{} emissivity showed excellent correspondence with hard X-ray pulses.

The time-scales of heating and cooling due to thermal conduction and radiation in the solar chromosphere are contingent upon various factors, including temperature, density, and specific physical processes involved. In the solar chromosphere, plasma has relatively low thermal conductivity, implying that heat conduction does not play a dominant role. Instead, other mechanisms, such as radiative cooling and plasma advection, have a greater influence on regulating the temperature and energy balance of the chromosphere. This introduces complex radiative transfer processes that depend on the temperature, density, and ionization state of the plasma.

Our study contributes to the ongoing exploration of solar flare ribbons and their intricate substructures. Earlier investigations have demonstrated that ribbons contain fine scale kernels and spectroscopic asymmetries \citep[e.g.,][]{dodson_flares_1956,harvey_flare_kernels_1971,Brannon2015,Janvier2016}.
 \citet{2024PietrowRibbon} showed that features underneath flare ribbons such as field concentrations, light bridges, and umbral dots, were frequently consistent with the locations of "freckles" visible on the flare ribbon. As a result of low polarization signatures in these regions and excess broadening of the H$\alpha$ line compared with other chromospheric lines, the authors interpreted these features as locations with greater column mass and higher line formation regions. The variety of atmospheric gradients due to underlying inhomogeneities have been demonstrated to impact flare ribbon formation and dynamics, and are thus a candidate for producing the observed variety of linear and non-linear riblet behaviours.
Several recent high-resolution studies provide independent context for the fine-scale ribbon-front structuring reported here. Using sub-arcsecond imaging, \citet{Faber2025} resolved fine-scale kernels and strands within flare ribbons, supporting the view that energy deposition is spatially intermittent at small spatial scales. Using SST/CHROMIS H$\beta$ spectroscopy, \citet{Faber2025b} reported narrow, vertically extended ribbon-front structures with typical downflow redshifts of order 16--21~km~s$^{-1}$, and interpreted these as ``riblets'' as defined in the present work. DKIST/ViSP spectropolarimetry further reports quasi-equally spaced ribbon-front ``blobs'' along Ca\,\textsc{ii}\,8542~\AA{} red-wing brightenings \citep{Yadav2025DKIST}, and high-cadence Balmer-continuum imaging reveals coherent quasi-harmonic pulsations co-spatial with rapidly evolving ribbon fronts \citep{song_spatiotemporal_2025}. Together, these results indicate that ribbon fronts can host coherent fine-scale substructure across multiple diagnostics and flaring conditions.

In lower cadence observations (15 s) of an X-class flare located on the solar disc, \citet{2024PietrowRibbon} found that the spectral profiles from the leading edges of the flare ribbons show Doppler red-shifted components that are within the variety of riblet speeds presented in this paper, and form a consistent analogue of the riblet features studied in our paper when viewed from a different angle \citep[see][Figures 7 and 8]{2024PietrowRibbon}. Furthermore, our observed riblet velocities align with the chromospheric condensation velocities with redshifts up to 17 km/s in a C-class flare \citep{keys_chromospheric_2011, li_simultaneous_2022} to downflows of up to 50 km/s in an X-class flare \citep{ashfield_iv_connecting_2022, graham_spectral_2020}. While the velocities we observe are similar to velocities in the \cite{druett_beam_2017} study on C-class flares, they fall considerably below the velocities expected from X-class flares, as indicated in \cite{druett_hydro2gen:_2018}. This discrepancy suggests that the role of the chromosphere might be more substantial than that of the electron beam in influencing the evolution of riblets. This finding underscores the necessity for further comprehensive investigations, involving the study of riblets across diverse flare scenarios. The advent of DKIST and its multiple Fabry-Perot spectral windows offer an opportunity to resolve the issue \citep{rast_dkist_2021}.

\section{Conclusion}\label{sec:conclusion}
High-cadence SST/CRISP observations resolve abundant fine structure within the flare ribbon. We refer to these coherent, thread-like substructures as riblets. Analogous features have also been described in the literature as ``jet-like'' structures. For the first time, their resolved time evolution allows a data driven kinematic classification into two categories: (i) Linear riblets that retract at approximately constant speed, and (ii) Non-linear riblets that exhibit measurable acceleration or deceleration.

A statistical comparison across key parameters (initial length, lifetime, average and maximum speeds, and start times) reveals no evidence for collective behaviour that differentiates these two populations. Spatially, their appearances are distributed along the entire ribbon rather than clustered at preferred locations and temporally, they occur throughout the rise phase without discernible sequencing between classes.

Multi-wavelength context also shows no robust interdependence: both classes occur across a broad range of AIA 1700\AA{} and 304\AA{} intensities and in proximity to RHESSI 25--50~keV contours, and the timing across UV, EUV, and X-ray diagnostics does not isolate a unique driver. Taken together, the absence of clear spatial, temporal, or wavelength linked distinctions means no definitive statement can yet be made about the physical processes that generate riblets. Future progress will require similarly high cadence imaging combined with spectroscopy and forward modelling across multiple flares to discriminate between competing scenarios.

\section{Acknowledgements}
All authors acknowledge UK Science and Technology Facilities Council (STFC) for IDL support. 
V.S. was supported by the STFC Northumbria 2021 DTP grant ST/W507696/1. 
NLSJ gratefully acknowledges the current financial support from STFC Grant ST/X001008/1. 
M.D. is supported by FWO project G0B4521N. AP is supported by the Deutsche Forschungsgemeinschaft,
DFG project number PI 2102/1-1. The authors wish to thank Irish Centre for High-End Computing (ICHEC) for the provision of computing facilities and support. We also like to thank the Solarnet project which is supported by the European Commission’s FP7 Capacities Programme under Grant Agreement number 312495 for T\&S. JGD would like to thank the Leverhulme Trust for an Emeritus Fellowship. Facilities: Swedish 1-meter Solar Telescope (SST)
The Swedish 1-meter Solar Telescope is operated on the island of La Palma by the Institute for Solar Physics of Stockholm University in the Spanish Observatorio del Roque de los Muchachos of the Instituto de Astrof\'isica de Canarias.
\newline
Software: CRISPEX
\newpage
\bibliography{main_bib}{}

@ARTICLE{kowalski_flare_model_obs_2017,
       author = {{Kowalski}, Adam F. and {Allred}, Joel C. and {Daw}, Adrian and {Cauzzi}, Gianna and {Carlsson}, Mats},
        title = "{The Atmospheric Response to High Nonthermal Electron Beam Fluxes in Solar Flares. I. Modeling the Brightest NUV Footpoints in the X1 Solar Flare of 2014 March 29}",
      journal = {\apj},
         year = 2017,
        month = feb,
       volume = {836},
       number = {1},
          eid = {12},
        pages = {12},
          doi = {10.3847/1538-4357/836/1/12},
archivePrefix = {arXiv},
       eprint = {1609.07390},
 primaryClass = {astro-ph.SR},
       adsurl = {https://ui.adsabs.harvard.edu/abs/2017ApJ...836...12K}
}

@ARTICLE{kowalski_radyn_2015,
       author = {{Kowalski}, Adam F. and {Hawley}, S.~L. and {Carlsson}, M. and {Allred}, J.~C. and {Uitenbroek}, H. and {Osten}, R.~A. and {Holman}, G.},
        title = "{New Insights into White-Light Flare Emission from Radiative-Hydrodynamic Modeling of a Chromospheric Condensation}",
      journal = {\solphys},
         year = 2015,
        month = dec,
       volume = {290},
       number = {12},
        pages = {3487-3523},
          doi = {10.1007/s11207-015-0708-x},
archivePrefix = {arXiv},
       eprint = {1503.07057},
 primaryClass = {astro-ph.SR},
       adsurl = {https://ui.adsabs.harvard.edu/abs/2015SoPh..290.3487K}
}

@ARTICLE{rast_dkist_2021,
       author = {{Rast}, Mark P. and {Bello Gonz{\'a}lez}, Nazaret and {Bellot Rubio}, Luis and {Cao}, Wenda and {Cauzzi}, Gianna and {Deluca}, Edward and {de Pontieu}, Bart and {Fletcher}, Lyndsay and {Gibson}, Sarah E. and {Judge}, Philip G. and {Katsukawa}, Yukio and {Kazachenko}, Maria D. and {Khomenko}, Elena and {Landi}, Enrico and {Mart{\'\i}nez Pillet}, Valent{\'\i}n and {Petrie}, Gordon J.~D. and {Qiu}, Jiong and {Rachmeler}, Laurel A. and {Rempel}, Matthias and {Schmidt}, Wolfgang and {Scullion}, Eamon and {Sun}, Xudong and {Welsch}, Brian T. and {Andretta}, Vincenzo and {Antolin}, Patrick and {Ayres}, Thomas R. and {Balasubramaniam}, K.~S. and {Ballai}, Istvan and {Berger}, Thomas E. and {Bradshaw}, Stephen J. and {Campbell}, Ryan J. and {Carlsson}, Mats and {Casini}, Roberto and {Centeno}, Rebecca and {Cranmer}, Steven R. and {Criscuoli}, Serena and {Deforest}, Craig and {Deng}, Yuanyong and {Erd{\'e}lyi}, Robertus and {Fedun}, Viktor and {Fischer}, Catherine E. and {Gonz{\'a}lez Manrique}, Sergio J. and {Hahn}, Michael and {Harra}, Louise and {Henriques}, Vasco M.~J. and {Hurlburt}, Neal E. and {Jaeggli}, Sarah and {Jafarzadeh}, Shahin and {Jain}, Rekha and {Jefferies}, Stuart M. and {Keys}, Peter H. and {Kowalski}, Adam F. and {Kuckein}, Christoph and {Kuhn}, Jeffrey R. and {Kuridze}, David and {Liu}, Jiajia and {Liu}, Wei and {Longcope}, Dana and {Mathioudakis}, Mihalis and {McAteer}, R.~T. James and {McIntosh}, Scott W. and {McKenzie}, David E. and {Miralles}, Mari Paz and {Morton}, Richard J. and {Muglach}, Karin and {Nelson}, Chris J. and {Panesar}, Navdeep K. and {Parenti}, Susanna and {Parnell}, Clare E. and {Poduval}, Bala and {Reardon}, Kevin P. and {Reep}, Jeffrey W. and {Schad}, Thomas A. and {Schmit}, Donald and {Sharma}, Rahul and {Socas-Navarro}, Hector and {Srivastava}, Abhishek K. and {Sterling}, Alphonse C. and {Suematsu}, Yoshinori and {Tarr}, Lucas A. and {Tiwari}, Sanjiv and {Tritschler}, Alexandra and {Verth}, Gary and {Vourlidas}, Angelos and {Wang}, Haimin and {Wang}, Yi-Ming and {NSO and DKIST Project} and {DKIST Instrument Scientists} and {DKIST Science Working Group} and {DKIST Critical Science Plan Community}},
        title = "{Critical Science Plan for the Daniel K. Inouye Solar Telescope (DKIST)}",
      journal = {\solphys},
         year = 2021,
        month = apr,
       volume = {296},
       number = {4},
          eid = {70},
        pages = {70},
          doi = {10.1007/s11207-021-01789-2},
archivePrefix = {arXiv},
       eprint = {2008.08203},
 primaryClass = {astro-ph.SR},
       adsurl = {https://ui.adsabs.harvard.edu/abs/2021SoPh..296...70R}
}

@ARTICLE{inchimoto_flare_obs_1984,
       author = {{Ichimoto}, K. and {Kurokawa}, H.},
        title = "{H{\ensuremath{\alpha}} Red Asymmetry of Solar Flares}",
      journal = {\solphys},
         year = 1984,
        month = jun,
       volume = {93},
       number = {1},
        pages = {105-121},
          doi = {10.1007/BF00156656},
       adsurl = {https://ui.adsabs.harvard.edu/abs/1984SoPh...93..105I}
}

@ARTICLE{osborne_flare_kernels_2022,
       author = {{Osborne}, Christopher M.~J. and {Fletcher}, Lyndsay},
        title = "{Flare kernels may be smaller than you think: modelling the radiative response of chromospheric plasma adjacent to a solar flare}",
      journal = {\mnras},
         year = 2022,
        month = nov,
       volume = {516},
       number = {4},
        pages = {6066-6074},
          doi = {10.1093/mnras/stac2570},
archivePrefix = {arXiv},
       eprint = {2209.03238},
 primaryClass = {astro-ph.SR},
       adsurl = {https://ui.adsabs.harvard.edu/abs/2022MNRAS.516.6066O}
}

@ARTICLE{harvey_flare_kernels_1971,
       author = {{Harvey}, Karen L.},
        title = "{The Explosive Phase of Solar Flares}",
      journal = {\solphys},
         year = 1971,
        month = feb,
       volume = {16},
       number = {2},
        pages = {423-430},
          doi = {10.1007/BF00162485},
       adsurl = {https://ui.adsabs.harvard.edu/abs/1971SoPh...16..423H}
}

@ARTICLE{dodson_flares_1956,
       author = {{Dodson}, Helen W. and {Hedeman}, E. Ruth and {McMath}, Robert R.},
        title = "{Photometry of Solar Flares.}",
      journal = {\apjs},
         year = 1956,
        month = feb,
       volume = {2},
        pages = {241},
          doi = {10.1086/190027},
       adsurl = {https://ui.adsabs.harvard.edu/abs/1956ApJS....2..241D}
}

@ARTICLE{kuridze_Halpha_asymmetries_2015,
       author = {{Kuridze}, D. and {Mathioudakis}, M. and {Sim{\~o}es}, P.~J.~A. and {Rouppe van der Voort}, L. and {Carlsson}, M. and {Jafarzadeh}, S. and {Allred}, J.~C. and {Kowalski}, A.~F. and {Kennedy}, M. and {Fletcher}, L. and {Graham}, D. and {Keenan}, F.~P.},
        title = "{H{\ensuremath{\alpha}} Line Profile Asymmetries and the Chromospheric Flare Velocity Field}",
      journal = {\apj},
         year = 2015,
        month = nov,
       volume = {813},
       number = {2},
          eid = {125},
        pages = {125},
          doi = {10.1088/0004-637X/813/2/125},
archivePrefix = {arXiv},
       eprint = {1510.01877},
 primaryClass = {astro-ph.SR},
       adsurl = {https://ui.adsabs.harvard.edu/abs/2015ApJ...813..125K}
}

@article{asai_evolution_2003,
	title = {Evolution of flare ribbons and energy release},
	volume = {32},
	issn = {0273-1177},
	url = {https://ui.adsabs.harvard.edu/abs/2003AdSpR..32.2561A},
	doi = {10.1016/j.asr.2003.03.008},
	urldate = {2023-01-12},
	journal = {Advances in Space Research},
	author = {Asai, A. and Masuda, S. and Yokoyama, T. and Shimojo, M. and Kurokawa, H. and Ishii, T. T. and Shibatal, K.},
	month = jan,
	year = {2003},
	note = {ADS Bibcode: 2003AdSpR..32.2561A},
	pages = {2561--2566},
}

@article{simoes_spectral_2019,
	title = {The {Spectral} {Content} of \textit{{SDO}} /{AIA} 1600 and 1700 Å {Filters} from {Flare} and {Plage} {Observations}},
	volume = {870},
	issn = {1538-4357},
	url = {https://iopscience.iop.org/article/10.3847/1538-4357/aaf28d},
	doi = {10.3847/1538-4357/aaf28d},
	number = {2},
	urldate = {2023-01-12},
	journal = {The Astrophysical Journal},
	author = {Simões, Paulo J. A. and Reid, Hamish A. S. and Milligan, Ryan O. and Fletcher, Lyndsay},
	month = jan,
	year = {2019},
	pages = {114},
}

@article{fletcher_observational_2011,
	title = {An {Observational} {Overview} of {Solar} {Flares}},
	volume = {159},
	issn = {0038-6308},
	url = {https://ui.adsabs.harvard.edu/abs/2011SSRv..159...19F},
	doi = {10.1007/s11214-010-9701-8},
	urldate = {2023-01-12},
	journal = {Space Science Reviews},
	author = {Fletcher, L. and Dennis, B. R. and Hudson, H. S. and Krucker, S. and Phillips, K. and Veronig, A. and Battaglia, M. and Bone, L. and Caspi, A. and Chen, Q. and Gallagher, P. and Grigis, P. T. and Ji, H. and Liu, W. and Milligan, R. O. and Temmer, M.},
	month = sep,
	year = {2011},
	note = {ADS Bibcode: 2011SSRv..159...19F},
	pages = {19--106},
}

@article{kazachenko_database_2017,
	title = {A {Database} of {Flare} {Ribbon} {Properties} from the \textit{{Solar} {Dynamics} {Observatory}} . {I}. {Reconnection} {Flux}},
	volume = {845},
	issn = {1538-4357},
	url = {https://iopscience.iop.org/article/10.3847/1538-4357/aa7ed6},
	doi = {10.3847/1538-4357/aa7ed6},
	number = {1},
	urldate = {2023-01-13},
	journal = {The Astrophysical Journal},
	author = {Kazachenko, Maria D. and Lynch, Benjamin J. and Welsch, Brian T. and Sun, Xudong},
	month = aug,
	year = {2017},
	pages = {49},
}

@article{allred_radiative_2005,
	title = {Radiative {Hydrodynamic} {Models} of the {Optical} and {Ultraviolet} {Emission} from {Solar} {Flares}},
	volume = {630},
	issn = {0004-637X, 1538-4357},
	url = {https://iopscience.iop.org/article/10.1086/431751},
	doi = {10.1086/431751},
	language = {en},
	number = {1},
	urldate = {2023-02-13},
	journal = {The Astrophysical Journal},
	author = {Allred, Joel C. and Hawley, Suzanne L. and Abbett, William P. and Carlsson, Mats},
	month = sep,
	year = {2005},
	pages = {573--586},
}

@article{druett_beam_2017,
	title = {Beam electrons as a source of {Hα} flare ribbons},
	volume = {8},
	issn = {2041-1723},
	url = {https://www.nature.com/articles/ncomms15905},
	doi = {10.1038/ncomms15905},
	language = {en},
	number = {1},
	urldate = {2023-02-13},
	journal = {Nature Communications},
	author = {Druett, Malcolm and Scullion, Eamon and Zharkova, Valentina and Matthews, Sarah and Zharkov, Sergei and Rouppe Van der Voort, Luc},
	month = jun,
	year = {2017},
	pages = {15905},
}

@article{druett_hydro2gen:_2018,
	title = {{HYDRO2GEN}: {Non}-thermal hydrogen {Balmer} and {Paschen} emission in solar flares generated by electron beams},
	volume = {610},
	issn = {0004-6361, 1432-0746},
	shorttitle = {{HYDRO2GEN}},
	url = {https://www.aanda.org/10.1051/0004-6361/201731053},
	doi = {10.1051/0004-6361/201731053},
	urldate = {2023-02-13},
	journal = {Astronomy \& Astrophysics},
	author = {Druett, M. K. and Zharkova, V. V.},
	month = feb,
	year = {2018},
	pages = {A68},
}

@article{scharmer_crisp_2008,
	title = {{CRISP} {Spectropolarimetric} {Imaging} of {Penumbral} {Fine} {Structure}},
	volume = {689},
	issn = {0004-637X, 1538-4357},
	url = {https://iopscience.iop.org/article/10.1086/595744},
	doi = {10.1086/595744},
	language = {en},
	number = {1},
	urldate = {2023-02-13},
	journal = {The Astrophysical Journal},
	author = {Scharmer, G. B. and Narayan, G. and Hillberg, T. and de la Cruz Rodriguez, J. and Löfdahl, M. G. and Kiselman, D. and Sütterlin, P. and van Noort, M. and Lagg, A.},
	month = dec,
	year = {2008},
	pages = {L69--L72},
}

@article{wuelser_high_1989,
	title = {High time resolution observations of {H} alpha line profiles during the impulsive phase of a solar flare},
	volume = {341},
	issn = {0004-637X, 1538-4357},
	url = {http://adsabs.harvard.edu/doi/10.1086/167567},
	doi = {10.1086/167567},
	language = {en},
	urldate = {2023-02-13},
	journal = {The Astrophysical Journal},
	author = {Wuelser, Jean-Pierre and Marti, Hans},
	month = jun,
	year = {1989},
	pages = {1088},
}

@inproceedings{sst,
author = {Goran B. Scharmer and Klas Bjelksjo and Tapio K. Korhonen and Bo Lindberg and Bertil Petterson},
title = {{The 1-m Swedish solar telescope}},
volume = {4853},
booktitle = {Innovative Telescopes and Instrumentation for Solar Astrophysics},
editor = {Stephen L. Keil and Sergey V. Avakyan},
organization = {International Society for Optics and Photonics},
publisher = {SPIE},
pages = {341 -- 350},
year = {2003},
doi = {10.1117/12.460377},
URL = {https://doi.org/10.1117/12.460377}
}

@ARTICLE{2023XuExtremeRedWing,
       author = {{Xu}, Yan and {Kerr}, Graham S. and {Polito}, Vanessa and {Huang}, Nengyi and {Jing}, Ju and {Wang}, Haimin},
        title = "{Extreme Red-wing Enhancements of UV Lines During the 2022 March 30 X1.3 Solar Flare}",
      journal = {arXiv e-prints},
         year = 2023,
        month = sep,
          eid = {arXiv:2309.05745},
        pages = {arXiv:2309.05745},
          doi = {10.48550/arXiv.2309.05745},
archivePrefix = {arXiv},
       eprint = {2309.05745},
 primaryClass = {astro-ph.SR},
       adsurl = {https://ui.adsabs.harvard.edu/abs/2023arXiv230905745X}
}

@INPROCEEDINGS{mats02,
       author = {{L{\"o}fdahl}, Mats G.},
        title = "{Multi-frame blind deconvolution with linear equality constraints}",
    booktitle = {Image Reconstruction from Incomplete Data II},
         year = 2002,
       editor = {{Bones}, Philip J. and {Fiddy}, Michael A. and {Millane}, Rick P.},
       series = {\procspie},
       volume = {4792},
        month = dec,
        pages = {146},
          doi = {10.1117/12.451791},
archivePrefix = {arXiv},
       eprint = {physics/0209004},
 primaryClass = {physics.optics},
       adsurl = {https://ui.adsabs.harvard.edu/abs/2002SPIE.4792..146L}
}

@ARTICLE{Neupert1968,
       author = {{Neupert}, Werner M.},
        title = "{Comparison of Solar X-Ray Line Emission with Microwave Emission during Flares}",
      journal = {\apjl},
         year = 1968,
        month = jul,
       volume = {153},
        pages = {L59},
          doi = {10.1086/180220},
       adsurl = {https://ui.adsabs.harvard.edu/abs/1968ApJ...153L..59N}
}

@article{2024PietrowRibbon,
	title = {Spectral variations within solar flare ribbons},
	volume = {685},
	copyright = {https://creativecommons.org/licenses/by/4.0},
	issn = {0004-6361, 1432-0746},
	url = {https://www.aanda.org/10.1051/0004-6361/202348839},
	doi = {10.1051/0004-6361/202348839},
	urldate = {2025-09-09},
	journal = {Astronomy \& Astrophysics},
	author = {Pietrow, A. G. M. and Druett, M. K. and Singh, V.},
	month = may,
	year = {2024},
	pages = {A137},
}

@article{vissers_flocculent_2012,
	title = {{FLOCCULENT} {FLOWS} {IN} {THE} {CHROMOSPHERIC} {CANOPY} {OF} {A} {SUNSPOT}},
	volume = {750},
	issn = {0004-637X, 1538-4357},
	url = {https://iopscience.iop.org/article/10.1088/0004-637X/750/1/22},
	doi = {10.1088/0004-637X/750/1/22},
	number = {1},
	urldate = {2023-06-19},
	journal = {The Astrophysical Journal},
	author = {Vissers, Gregal and Rouppe Van Der Voort, Luc},
	month = may,
	year = {2012},
	pages = {22},
}

@article{carrington_description_1859,
	title = {Description of a {Singular} {Appearance} seen in the {Sun} on {September} 1, 1859},
	volume = {20},
	issn = {0035-8711, 1365-2966},
	url = {https://academic.oup.com/mnras/article-lookup/doi/10.1093/mnras/20.1.13},
	doi = {10.1093/mnras/20.1.13},
	language = {en},
	number = {1},
	urldate = {2023-06-29},
	journal = {Monthly Notices of the Royal Astronomical Society},
	author = {Carrington, R. C.},
	month = nov,
	year = {1859},
	pages = {13--15},
}

@article{lin_reuven_2002,
	title = {The {Reuven} {Ramaty} {High}-{Energy} {Solar} {Spectroscopic} {Imager} ({RHESSI})},
	volume = {210},
	issn = {1573-093X},
	url = {https://doi.org/10.1023/A:1022428818870},
	doi = {10.1023/A:1022428818870},
	language = {en},
	number = {1},
	urldate = {2023-07-17},
	journal = {Solar Physics},
	author = {Lin, R.P. and Dennis, B.R. and Hurford, G.J. and Smith, D.M. and Zehnder, A. and Harvey, P.R. and Curtis, D.W. and Pankow, D. and Turin, P. and Bester, M. and Csillaghy, A. and Lewis, M. and Madden, N. and van Beek, H.F. and Appleby, M. and Raudorf, T. and McTiernan, J. and Ramaty, R. and Schmahl, E. and Schwartz, R. and Krucker, S. and Abiad, R. and Quinn, T. and Berg, P. and Hashii, M. and Sterling, R. and Jackson, R. and Pratt, R. and Campbell, R.D. and Malone, D. and Landis, D. and Barrington-Leigh, C.P. and Slassi-Sennou, S. and Cork, C. and Clark, D. and Amato, D. and Orwig, L. and Boyle, R. and Banks, I.S. and Shirey, K. and Tolbert, A.K. and Zarro, D. and Snow, F. and Thomsen, K. and Henneck, R. and Mchedlishvili, A. and Ming, P. and Fivian, M. and Jordan, John and Wanner, Richard and Crubb, Jerry and Preble, J. and Matranga, M. and Benz, A. and Hudson, H. and Canfield, R.C. and Holman, G.D. and Crannell, C. and Kosugi, T. and Emslie, A.G. and Vilmer, N. and Brown, J.C. and Johns-Krull, C. and Aschwanden, M. and Metcalf, T. and Conway, A.},
	month = nov,
	year = {2002},
	pages = {3--32},
}

@article{meegan_fermi_2009,
	title = {{THE} \textit{{FERMI}} {GAMMA}-{RAY} {BURST} {MONITOR}},
	volume = {702},
	issn = {0004-637X, 1538-4357},
	url = {https://iopscience.iop.org/article/10.1088/0004-637X/702/1/791},
	doi = {10.1088/0004-637X/702/1/791},
	number = {1},
	urldate = {2023-07-17},
	journal = {The Astrophysical Journal},
	author = {Meegan, Charles and Lichti, Giselher and Bhat, P. N. and Bissaldi, Elisabetta and Briggs, Michael S. and Connaughton, Valerie and Diehl, Roland and Fishman, Gerald and Greiner, Jochen and Hoover, Andrew S. and Van Der Horst, Alexander J. and Von Kienlin, Andreas and Kippen, R. Marc and Kouveliotou, Chryssa and McBreen, Sheila and Paciesas, W. S. and Preece, Robert and Steinle, Helmut and Wallace, Mark S. and Wilson, Robert B. and Wilson-Hodge, Colleen},
	month = sep,
	year = {2009},
	pages = {791--804},
}

@article{fisher_dynamics_1989,
	title = {Dynamics of flare-driven chromospheric condensations},
	volume = {346},
	issn = {0004-637X, 1538-4357},
	url = {http://adsabs.harvard.edu/doi/10.1086/168084},
	doi = {10.1086/168084},
	language = {en},
	urldate = {2023-07-17},
	journal = {The Astrophysical Journal},
	author = {Fisher, George H.},
	month = nov,
	year = {1989},
	pages = {1019},
}

@article{abbett_dynamic_1999,
	title = {Dynamic {Models} of {Optical} {Emission} in {Impulsive} {Solar} {Flares}},
	volume = {521},
	issn = {0004-637X, 1538-4357},
	url = {https://iopscience.iop.org/article/10.1086/307576},
	doi = {10.1086/307576},
	language = {en},
	number = {2},
	urldate = {2023-07-17},
	journal = {The Astrophysical Journal},
	author = {Abbett, William P. and Hawley, Suzanne L.},
	month = aug,
	year = {1999},
	pages = {906--919},
}

@article{ashfield_iv_connecting_2022,
	title = {Connecting {Chromospheric} {Condensation} {Signatures} to {Reconnection}-driven {Heating} {Rates} in an {Observed} {Flare}},
	volume = {926},
	issn = {0004-637X, 1538-4357},
	url = {https://iopscience.iop.org/article/10.3847/1538-4357/ac402d},
	doi = {10.3847/1538-4357/ac402d},
	number = {2},
	urldate = {2023-07-17},
	journal = {The Astrophysical Journal},
	author = {Ashfield Iv, William H. and Longcope, Dana W. and Zhu, Chunming and Qiu, Jiong},
	month = feb,
	year = {2022},
	pages = {164},
}

@article{li_simultaneous_2022,
	title = {Simultaneous {Observations} of {Chromospheric} {Evaporation} and {Condensation} during a {C}-class {Flare}},
	volume = {926},
	issn = {0004-637X, 1538-4357},
	url = {https://iopscience.iop.org/article/10.3847/1538-4357/ac426b},
	doi = {10.3847/1538-4357/ac426b},
	number = {1},
	urldate = {2023-07-17},
	journal = {The Astrophysical Journal},
	author = {Li, Dong and Hong, Zhenxiang and Ning, Zongjun},
	month = feb,
	year = {2022},
	pages = {23},
}

@ARTICLE{2005SoPh..228..191V,
       author = {{van Noort}, Michiel and {Rouppe van der Voort}, Luc and
         {L{\"o}fdahl}, Mats G.},
        title = "{Solar Image Restoration By Use Of Multi-frame Blind De-convolution With Multiple Objects And Phase Diversity}",
      journal = {\solphys},
         year = 2005,
        month = may,
       volume = {228},
       number = {1-2},
        pages = {191-215},
          doi = {10.1007/s11207-005-5782-z},
       adsurl = {https://ui.adsabs.harvard.edu/abs/2005SoPh..228..191V}
}

@article{falchi_chromospheric_2002,
	title = {Chromospheric models of a solar flare including velocity fields},
	volume = {387},
	issn = {0004-6361, 1432-0746},
	url = {http://www.aanda.org/10.1051/0004-6361:20020454},
	doi = {10.1051/0004-6361:20020454},
	number = {2},
	urldate = {2023-08-09},
	journal = {Astronomy \& Astrophysics},
	author = {Falchi, A. and Mauas, P. J. D.},
	month = may,
	year = {2002},
	pages = {678--686},
}

@article{benz_flare_2017,
	title = {Flare {Observations}},
	volume = {14},
	issn = {2367-3648, 1614-4961},
	url = {http://link.springer.com/10.1007/s41116-016-0004-3},
	doi = {10.1007/s41116-016-0004-3},
	language = {en},
	number = {1},
	urldate = {2023-08-09},
	journal = {Living Reviews in Solar Physics},
	author = {Benz, Arnold O.},
	month = dec,
	year = {2017},
	pages = {2},
}

@article{brown_solar_1980,
	title = {Solar flares},
	volume = {43},
	issn = {0034-4885, 1361-6633},
	url = {https://iopscience.iop.org/article/10.1088/0034-4885/43/2/001},
	doi = {10.1088/0034-4885/43/2/001},
	number = {2},
	urldate = {2023-08-09},
	journal = {Reports on Progress in Physics},
	author = {Brown, J C and Smith, D F},
	month = feb,
	year = {1980},
	pages = {125--197},
}

@article{doyle_diagnosing_2013,
	title = {Diagnosing transient ionization in dynamic events},
	volume = {557},
	issn = {0004-6361, 1432-0746},
	url = {http://www.aanda.org/10.1051/0004-6361/201321902},
	doi = {10.1051/0004-6361/201321902},
	urldate = {2023-10-12},
	journal = {Astronomy \& Astrophysics},
	author = {Doyle, J. G. and Giunta, A. and Madjarska, M. S. and Summers, H. and O’Mullane, M. and Singh, A.},
	month = sep,
	year = {2013},
	pages = {L9},
}

@article{doyle_diagnostic_2012,
	title = {The {Diagnostic} {Potential} of {Transition} {Region} {Lines} {Undergoing} {Transient} {Ionization} in {Dynamic} {Events}},
	volume = {280},
	issn = {0038-0938, 1573-093X},
	url = {http://link.springer.com/10.1007/s11207-012-0025-6},
	doi = {10.1007/s11207-012-0025-6},
	language = {en},
	number = {1},
	urldate = {2023-10-12},
	journal = {Solar Physics},
	author = {Doyle, J. G. and Giunta, A. and Singh, A. and Madjarska, M. S. and Summers, H. and Kellett, B. J. and O’Mullane, M.},
	month = sep,
	year = {2012},
	pages = {111--124},
}

@article{keys_chromospheric_2011,
	title = {Chromospheric velocities of a {C}-class flare},
	volume = {529},
	issn = {0004-6361, 1432-0746},
	url = {http://www.aanda.org/10.1051/0004-6361/201016320},
	doi = {10.1051/0004-6361/201016320},
	urldate = {2023-10-24},
	journal = {Astronomy \& Astrophysics},
	author = {Keys, P. H. and Jess, D. B. and Mathioudakis, M. and Keenan, F. P.},
	month = may,
	year = {2011},
	pages = {A127},
}

@article{graham_spectral_2020,
	title = {Spectral {Signatures} of {Chromospheric} {Condensation} in a {Major} {Solar} {Flare}},
	volume = {895},
	issn = {1538-4357},
	url = {https://iopscience.iop.org/article/10.3847/1538-4357/ab88ad},
	doi = {10.3847/1538-4357/ab88ad},
	number = {1},
	urldate = {2023-10-24},
	journal = {The Astrophysical Journal},
	author = {Graham, David R. and Cauzzi, Gianna and Zangrilli, Luca and Kowalski, Adam and Simões, Paulo and Allred, Joel},
	month = may,
	year = {2020},
	pages = {6},
}

@article{de_la_cruz_rodriguez_crispred:_2015,
	title = {{CRISPRED}: {A} data pipeline for the {CRISP} imaging spectropolarimeter},
	volume = {573},
	issn = {0004-6361, 1432-0746},
	shorttitle = {{CRISPRED}},
	url = {http://www.aanda.org/10.1051/0004-6361/201424319},
	doi = {10.1051/0004-6361/201424319},
	urldate = {2023-11-02},
	journal = {Astronomy \& Astrophysics},
	author = {de la Cruz Rodríguez, J. and Löfdahl, M. G. and Sütterlin, P. and Hillberg, T. and Rouppe Van Der Voort, L.},
	month = jan,
	year = {2015},
	pages = {A40},
}

@article{Faber2025,
  author  = {Faber, Jonas Thoen and Joshi, Reetika and Rouppe van der Voort, Luc and Wedemeyer, Sven and Fletcher, Lyndsay and Aulanier, Guillaume and N\'obrega-Siverio, Daniel},
  title   = {High-resolution observational analysis of flare ribbon fine structures},
  journal = {Astronomy \& Astrophysics},
  volume  = {693},
  pages   = {A8},
  year    = {2025},
  doi     = {10.1051/0004-6361/202452370},
}

@article{Li2019,
  author  = {Li, Xiaohong and Zhang, Jun and Yang, Shuhong and Hou, Yijun},
  title   = {Solar jet-like features rooted in flare ribbons},
  journal = {Publications of the Astronomical Society of Japan},
  volume  = {71},
  pages   = {14},
  year    = {2019},
  doi     = {10.1093/pasj/psy128},
}

@article{Brannon2015,
  author = {Brannon, S. R. and Longcope, D. W. and Qiu, J.},
  title = {Spectroscopic observations of evolving flare ribbon substructure suggesting origin in current sheet waves},
  journal = {Astrophysical Journal},
  volume = {810},
  number = {1},
  pages = {4},
  year = {2015},
  doi = {10.1088/0004-637X/810/1/4}
}

@article{Janvier2016,
  author = {Janvier, M. and Savcheva, A. and Pariat, E. and Tassev, S. and Millholland, S. and Bommier, V. and McCauley, P. and McKillop, S. and Dougan, F.},
  title = {Evolution of flare ribbons, electric currents and quasi-separatrix layers during an X-class flare},
  journal = {Astronomy \& Astrophysics},
  volume = {591},
  pages = {A141},
  year = {2016},
  doi = {10.1051/0004-6361/201628406}
}

@ARTICLE{song_spatiotemporal_2025,
  author       = {{Song}, D.-C. and {Dominique}, M. and {Zimovets}, I. and et~al.},
  title        = "{Unveiling Spatiotemporal Properties of the Quasi-periodic Pulsations in the Balmer Continuum at 3600\,\AA\,in an X-class Solar White-light Flare}",
  journal      = {\apjl},
  year         = 2025,
  volume       = 983,
  pages        = {L41},
  doi          = {10.3847/2041-8213/adc4e9},
}

@article{Krucker2011,
  author  = {Krucker, S. and Hudson, H. S. and Jeffrey, N. L. S. and Battaglia, M. and Kontar, E. P. and Benz, A. O. and Csillaghy, A. and Lin, R. P.},
  title   = {High-resolution Imaging of Solar Flare Ribbons and Its Implication on the Thick-Target Beam Model},
  journal = {The Astrophysical Journal},
  year    = {2011},
  volume  = {739},
  pages   = {96},
  doi     = {10.1088/0004-637X/739/2/96}
}

@article{Kuhar2016,
  author  = {Kuhar, M. and Krucker, S. and Mart{\'\i}nez Oliveros, J. C. and Battaglia, M. and Kleint, L. and Casadei, D. and Hudson, H. S.},
  title   = {Correlation of Hard X-Ray and White Light Emission in Solar Flares},
  journal = {The Astrophysical Journal},
  year    = {2016},
  volume  = {816},
  pages   = {6},
  doi     = {10.3847/0004-637X/816/1/6}
}

@article{Yadav2025DKIST,
  author       = {Yadav, Rahul and Kazachenko, Maria D. and Cauzzi, Gianna and Tamburri, Cole A. and Corchado, Marcel and French, Ryan},
  title        = {Multi-line Spectropolarimetric Observation of Flare Ribbon Fine Structures with ViSP/DKIST},
  journal      = {arXiv e-prints},
  year         = {2025},
  eprint       = {2507.20070},
  archivePrefix= {arXiv},
  primaryClass = {astro-ph.SR},
  month        = {July}
}

@article{lofdahl_sstred:_2021,
	title = {{SSTRED}: {Data}- and metadata-processing pipeline for {CHROMIS} and {CRISP}},
	volume = {653},
	issn = {0004-6361, 1432-0746},
	shorttitle = {{SSTRED}},
	url = {https://www.aanda.org/10.1051/0004-6361/202141326},
	doi = {10.1051/0004-6361/202141326},
	urldate = {2023-06-19},
	journal = {Astronomy \& Astrophysics},
	author = {Löfdahl, Mats G. and Hillberg, Tomas and De La Cruz Rodríguez, Jaime and Vissers, Gregal and Andriienko, Oleksii and Scharmer, Göran B. and Haugan, Stein V. H. and Fredvik, Terje},
	month = sep,
	year = {2021},
	pages = {A68},
}

@article{Kennedy2015,
  author    = {Kennedy, M. B. and Milligan, R. O. and Allred, J. C. and Mathioudakis, M. and Keenan, F. P.},
  title     = {Radiative hydrodynamic modelling and observations of the X-class solar flare on 2011 March 9},
  journal   = {Astronomy \& Astrophysics},
  volume    = {578},
  pages     = {A72},
  year      = {2015},
  doi       = {10.1051/0004-6361/201425144}
}

@article{Allred2015,
  author    = {Allred, Joel C. and Kowalski, Adam F. and Carlsson, Mats},
  title     = {A Unified Computational Model for Solar and Stellar Flares},
  journal   = {The Astrophysical Journal},
  volume    = {809},
  number    = {1},
  pages     = {104},
  year      = {2015},
  doi       = {10.1088/0004-637X/809/1/104}
}

@article{Simoes2015,
  author  = {Sim{\~o}es, Paulo J. A. and Graham, David R. and Fletcher, Lyndsay},
  title   = {Impulsive Heating of Solar Flare Ribbons Above 10 MK},
  journal = {Solar Physics},
  volume  = {290},
  number  = {12},
  pages   = {3573--3591},
  year    = {2015},
  doi     = {10.1007/s11207-015-0709-9}
}

@article{James2023,
  author    = {James, T. and Reid, H. and Kontar, E. P. and Gordovskyy, M. and Jeffrey, N. L. S. and others},
  title     = {Statistical study of type III bursts and associated HXR emission: number of electrons from RHESSI and Fermi/GBM},
  journal   = {Astronomy \& Astrophysics},
  volume    = {674},
  pages     = {A129},
  year      = {2023},
  doi       = {10.1051/0004-6361/202245825}
}

@misc{Faber2025b,
	title = {Fine details in solar flare ribbons: {Statistical} insights from observations with the {Swedish} 1-m {Solar} {Telescope}},
	copyright = {Creative Commons Attribution 4.0 International},
	shorttitle = {Fine details in solar flare ribbons},
	url = {https://arxiv.org/abs/2510.23246},
	doi = {10.48550/ARXIV.2510.23246},
	urldate = {2026-01-15},
	publisher = {arXiv},
	author = {Faber, Jonas Thoen and Joshi, Reetika and van der Voort, Luc Rouppe and Wedemeyer, Sven and Øyre, Eilif Sommer and Poquet, Ignasi J. Soler and Brunvoll},
	year = {2025},
}

@article{kerr_spatial_variation_2026,
  title   = {Spatial variation of energy transport mechanisms within solar flare ribbons},
  author  = {Kerr, Graham S. and Krucker, S{\"a}m and Allred, Joel C. and Rodr{\'i}guez-G{\'o}mez, Jenny M. and Inglis, Andrew R. and Ryan, Daniel F. and Hayes, Laura A. and Milligan, Ryan O. and Kowalski, Adam F. and Plowman, Joseph E. and Young, Peter R. and Kucera, Therese A. and Brosius, Jeffrey W.},
  journal = {Nature Astronomy},
  year    = {2026},
  doi     = {10.1038/s41550-025-02747-9}
}

@article{sellers2022,
  author  = {Sellers, S. G. and Milligan, R. O. and McAteer, R. T. J.},
  title   = {Call and Response: A Time-resolved Study of Chromospheric Evaporation in a Large Solar Flare},
  journal = {The Astrophysical Journal},
  year    = {2022},
  volume  = {936},
  number  = {1},
  pages   = {85},
  doi     = {10.3847/1538-4357/ac87a9}
}

@article{fisher1985b,
  author  = {Fisher, G. H. and Canfield, R. C. and McClymont, A. N.},
  title   = {Flare Loop Radiative Hydrodynamics. {VI}. Chromospheric Evaporation due to Heating by Nonthermal Electrons},
  journal = {The Astrophysical Journal},
  year    = {1985},
  volume  = {289},
  pages   = {425--434},
  doi     = {10.1086/162902}
}
\bibliographystyle{aasjournal}

\end{document}